\RequirePackage{fix-cm}
\documentclass{article}
\usepackage{arxiv}
\usepackage{lipsum}
\usepackage{placeins}
\usepackage[italicComments=true,indLines=true,noEnd=false]{algpseudocodex}
\usepackage{float}
\usepackage[english]{babel}
\usepackage[utf8]{inputenc} % allow utf-8 input
\usepackage[T1]{fontenc}    % use 8-bit T1 fonts
\usepackage{hyperref}       % hyperlinks
\usepackage{url}            % simple URL typesetting
\usepackage{booktabs}       % professional-quality tables
\usepackage{stackrel,amssymb}
\usepackage{amsfonts,amsmath,amstext,amsbsy,amsthm}
\usepackage{nicefrac}       % compact symbols for 1/2, etc.
\usepackage[version=4]{mhchem}
\usepackage[nopatch=eqnum]{microtype}      % microtypography
\usepackage{graphicx}
\usepackage[numbers,sort&compress]{natbib}
\usepackage{doi}

\usepackage{xcolor}
\usepackage{enumitem}
\usepackage{caption}
\usepackage[title]{appendix}
\usepackage[capitalise,nameinlink]{cleveref}

\usepackage[bottom]{footmisc}
\usepackage{multirow}
\usepackage{subfigure}
\usepackage{pgfplots}
\usepackage{tikz}

\usepackage{algorithm}
\usepackage{algpseudocodex}

\newcounter{savesecnumdepth}
\providecommand{\backsectionname}{Acknowledgements}
\newcommand{\backsection}[2][\backsectionname]{\begingroup\par%
  \small%
  \setcounter{savesecnumdepth}{\value{secnumdepth}}%
  \setcounter{secnumdepth}{0}%
\vskip6pt
\noindent \textbf{#1.} #2\par%
  \setcounter{secnumdepth}{\value{savesecnumdepth}}%
  \endgroup}
\title{Learning the latent dynamics of Fluid flows from High-Fidelity Numerical Simulations using Parsimonious  Diffusion Maps}

\author{
\textbf{Alessandro Della Pia\textcolor{blue}{$^{1}$}, Dimitris Patsatzis\textcolor{blue}{$^{1}$}, Lucia Russo \textcolor{blue}{$^{2,*}$}, Constantinos Siettos\textcolor{blue}{$^{3,}$}\thanks{Corresponding authors, emails: \texttt{lucia.russo@stems.cnr.it, constantinos.siettos@unina.it}}}
{}\\
\textcolor{blue}{$^{(1)}$}Modelling Engineering Risk and Complexity, \emph{Scuola Superiore Meridionale}, Naples 80138, Italy \\
\textcolor{blue}{$^{(2)}$}Institute of Science and Technology for Energy and Sustainable Mobility, \emph{Consiglio Nazionale}\\ \emph{ delle Ricerche}, Naples 80125, Italy\\
\textcolor{blue}{$^{(3)}$}Dipartimento di Matematica e Applicazioni ‘‘Renato Caccioppoli", \emph{Universit\`a degli Studi di Napoli}\\ \emph{Federico II}, Naples 80126, Italy\\
}
\pgfplotsset{compat=1.18}

\begin{document}
\maketitle
\begin{abstract}
We use parsimonious diffusion maps (PDMs) to discover the latent dynamics of high-fidelity Navier-Stokes simulations with a focus on the 2D fluidic pinball problem. By varying the Reynolds number, different flow regimes emerge, ranging from steady symmetric flows to quasi-periodic asymmetric and  turbulence. We show, that the proposed non-linear manifold learning scheme, identifies in a crisp manner the expected intrinsic dimension of the underlying emerging dynamics over the parameter space. In particular, PDMs, estimate that the emergent dynamics in the oscillatory regime can be captured by just two variables, while in the chaotic regime, the dominant modes are three as anticipated by the normal form theory. On the other hand, proper orthogonal decomposition (POD)/PCA, most commonly used for dimensionality reduction in fluid mechanics, does not provide such a crisp separation between the dominant modes. To validate the performance of PDMs, we also computed the reconstruction error, by constructing a decoder using Geometric Harmonics. We show that the proposed scheme outperforms the POD/PCA over the whole Reynolds number range. Thus, we believe that the proposed scheme will allow for the development of more accurate reduced order models for high-fidelity fluid dynamics simulators, thus relaxing the curse of dimensionality in numerical analysis tasks such as bifurcation analysis, optimization and control.
%Five different flow regimes are considered, spanning from steady symmetric ($Re < 18$) to fully chaotic ($Re > 115$) conditions. In the first step of the Diffusion Maps embedding, the minimum set of DMs reduced coordinates (eigenvectors) necessary to represent the flow dynamics in all the regimes is found by projecting the high-dimensional simulation data into the reduced low-dimensional space (restriction operation).
%The nonlinear manifold lying in the state-space spanned by the three leading DMs coordinates is thus obtained by varying the Reynolds number $Re$, and its shape discussed with reference to the different physical mechanisms at play across the flow regimes. Then, the time series embedded into the manifold are lifted back to the original space by means of Geometric Harmonics (lifting operation), such as to evaluate the reconstruction error between the ``ground truth'' solution (i.e. high-fidelity simulation data) and the DMs low-dimensional ``reconstruction''. 
%The performance of the DMs-based reconstruction is finally compared with a counterpart linear technique based on the Proper Orthogonal Decomposition (POD), which demonstrates the superiority of the proposed approach in parsimoniously representing the nonlinear dynamics over the whole range of the $Re$.
%To the authors' knowledge, this is the first application of Diffusion Maps to discover parsimonious nonlinear coordinates of a two-dimensional flow configuration such as the fluidic pinball, which is characterized by a rich variety of flow regimes including chaos.
\end{abstract}

% keywords can be removed

\section{Introduction}
\label{sec:intro}
The Lorenz equations \cite{lorenz1963deterministic,cvitanovic2017universality} constitute a totem in reduced order modeling (ROM), nonlinear dynamics and chaos theory \cite{cvitanovic2005chaos}, parsimoniously capturing the essential dynamics in the solutions to the Navier-Stokes equations under certain conditions. Today, both theoretical and computational advances have significantly propelled the interest for integrating machine learning (ML) for learning ROMs from high-fidelity computational fluid dynamics (CFD) simulators \cite{brunton2020machine,raissi2020hidden,vinuesa_brunton_ML-CFD,mendez2023data}. %that may provide a low-dimensional but insightful representation of the flow. ROMs describe the evolution of such structures, thus providing a low dimensional characterization of the flow. Within the framework of ML, the construction of ROMs, involves the identification of a low-dimensional (based on the variables found in the previous step) surrogate models using for example 
Various methods have been proposed for the construction of surrogate ML models including Sparse identification of nonlinear dynamical systems (SINDy) \cite{brunton2016discovering}, Gaussian process regression (GPR) \cite{wan2017reduced,lee2020coarse,papaioannou2022time}, self-organizing maps \cite{alexandridis2002modelling} and deep-feedforward neural networks \cite{bertalan2019learning,lee2020coarse,arbabi2020linking,galaris2022numerical,lee2023learning,Kevrekidis_DM,dietrich2023learning,fabiani2024task}.
For dealing with the ``curse of dimensionality'' in the training phase of the ML surrogate models, a first step towards ROMs, usually involves the estimation of the intrinsic dimension of the latent space and a set of vectors that span them. In general, in the framework of data-driven methods, this can be achieved by the use of linear, such as the proper orthogonal decomposition (POD)/principal component analysis (PCA) \cite{deane1991low,rowley2005model,ma2002low,hijazi2020data,brunton2020machine} and nonlinear manifold learning algorithms such as diffusion maps \cite{coifman2005geometric,coifman2006geometric,nadler2006diffusion,dsilva2018parsimonious,galaris2022numerical,Kevrekidis_DM}. Autoencoders have been also used \cite{li2020scalable,vlachas2022multiscale,floryan2022data,Eivazi} for this task.
However, while POD/PCA has been extensively used in CFD for reducing high-dimensional flow data by projecting them onto a lower-dimensional linear subspace spanned by orthogonal modes, the application of nonlinear manifold learning techniques in this field remains relatively under-explored. As a result, while POD/PCA provides a simple approach due to the linear transformations for dimensionality reduction and data reconstruction in the high-dimensional space, it may miss some of the complexities that nonlinear manifold learning techniques could potentially reveal. %DMs have been recently used in \cite{Kevrekidis_DM} to build predictive models of a vertically falling liquid film flow configuration in case of missing data. 

Here,  we aim to assess the performance of PDMs, for discovering the intrinsic dynamics of high-fidelity Navier-Stokes simulations of the fluidic pinball configuration and compare it with the traditional POD/PCA approach via the reconstruction error in the high-dimensional space. The latter task is trivial for POD/PCA; for nonlinear manifold learning algorithms the solution of the pre-image problem (i.e. ``lifting'' the solutions in the latent space back to the original high-dimensional space)  is not unique, thus far from trivial \cite{chiavazzo2014reduced,papaioannou2022time}. 
The first low-order model of the fluidic pinball was developed by \cite{Deng_Noack_2020} by means of POD/Galerkin projection of Navier-Stokes equations. The ROM is build in two distinct flow regimes, respectively characterized by an Andronov-Hopf ($Re_1 \approx 18$) and a pitchfork ($Re_2 \approx 68$) bifurcation. The first Reynolds number threshold ($Re_1$) characterizes the  transition from a steady symmetric flow to a periodic symmetric vortex shedding. For $Re > Re_2$, the symmetry of the vortex shedding vanishes, and the flow enters into a periodic asymmetric regime. The flow pattern is characterized by an asymmetric vortex shedding with a base-bleeding jet in the near wake of the cylinders, which deflects upward or downward depending on which of the two stable limit cycles is reached. 
The resulting ROM has three degrees of freedom in the primary flow regime, five in the secondary one, and was able to capture well the dynamics resulting from the two successive bifurcations. Among the open questions highlighted in \cite{Deng_Noack_2020}, there was the possibility to extend the ROM range of applicability to higher $Re$ values. In particular, it was pointed out that the next flow transition by increasing $Re$ is a Neimark–Säcker bifurcation (for $Re_3 \approx 104$), which is characterized by a new oscillatory phenomenon related to the base bleeding jet. Therefore, it is expected that the number of degrees of freedom needed for the ROM to accurately describe the dynamics may increase in this new regime. Moreover, in the same work. the authors indicated as a future research goal the use of machine learning to learn latent spaces, modes and dynamical systems. Following this perspective, the same authors presented a Galerkin force model (\cite{Deng_Noack_2021}) and a self-supervised cluster-based hierarchical reduced-order modelling methodology (\cite{Deng_Noack_2022}) to model the complex dynamics of the fluidic pinball at higher Reynolds number values. In the last work, they proposed a cluster-based hierarchical network model able to identify the transient and post-transient dynamics characterizing the fluidic pinball for $Re > Re_3$ and $Re > Re_4$ ($Re_4 \approx 115$), where the flow exhibits chaotic dynamics. Later on, in \cite{Farzamnik_2023} it was used a nonlinear manifold learning technique to identify latent spaces from snapshot data. In particular, the authors used Isomap \cite{tenenbaum2000global} as encoder and K-nearest neighbours (KNN) algorithm as decoder. The proposed Isomap-KNN approach correctly identified the periodic asymmetric regime at $Re=80$ resulting from the pitchfork bifurcation, and the chaotic regime at $Re=130$. Their identified latent space had three degrees of freedom, which are linked to physical mechanisms at play in the flow. 

Here, we first performed two-dimensional numerical simulations of the incompressible Navier–Stokes equations to compute the viscous wake flow behind the fluidic pinball by varying the Reynolds number $Re$. Five different flow regimes emerge, ranging from steady symmetric flow ($Re < 18$) to spatio-temporal chaos ($Re > 115$). Then, we employed PDMs \cite{dsilva2018parsimonious,galaris2022numerical,gallos2024data} to represent the flow dynamics in latent spaces in all regimes. %The nonlinear manifold lying in the state-space spanned by the three leading DMs coordinates is thus obtained by varying the Reynolds number $Re$, and its shape discussed with reference to the different physical mechanisms at play across the flow regimes. 
Then, to compute the reconstruction error between the ``ground truth'' solution (i.e. high-fidelity simulations) and the PDMs low-dimensional ``reconstruction'', we solved the pre-image problem with the aid of Geometric Harmonics \cite{coifman2006geometric,chiavazzo2014reduced,evangelou2022double,papaioannou2022time,Patsatzis_2023,gallos2024data}.
The performance of the PDMs-based reconstruction is finally compared with the traditional POD/PCA approach. %To the authors' knowledge, this is the first application of DMs to discover parsimonious nonlinear coordinates of a high-fidelity two-dimensional flow configuration of the fluidic pinball, and its comparison to the POD/PCA.

The rest of the work is organized as follows. In section ~\ref{sec:layout}, we describe the fluidic pinball configuration, focusing on the numerical methodology employed to obtain the data and discussing the different emerging flow regimes. In section ~\ref{sec:methods}, we describe the parsimonious PDM algorithm, together with the restriction (based on the Nyström extension) and lifting (based on Geometric Harmonics) operations necessary to project the time series embedded into the manifold back to the original space. Results are discussed in section ~\ref{sec:results}, and conclusions are finally provided in section ~\ref{sec:conclusions}.

\section{The fluidic pinball problem}
\label{sec:layout}
For our illustration, we focus on the fluidic pinball that has been shown to be a suitable test-bed configuration to study general flow phenomena like bifurcations and flow control (\cite{Deng_Noack_2020}).
The wake flow characterizing this configuration undergoes a set of interesting transitions at different values of the Reynolds number $Re$, which measures the relative influence between inertia and viscous forces within the flow. As a matter of fact, this allows exploration of reduced order modelling and flow control strategies in a wide range of scenarios. 

%However, it was shown that the residual variance of embedding procedure at $Re=130$ was higher than at lower $Re$. It is therefore reasonable to expect that, as $Re$ increases and more chaotic regimes are entered, the true dimensionality is higher due to the arising of a more complex dynamics. Therefore, more degrees of freedom may be necessary to accurately describe the flow dynamics at higher $Re$. In other words, the dimension of the latent space is expected to increase with $Re$. 
\begin{figure}%[h]
	\centering
	\includegraphics[scale=.8]{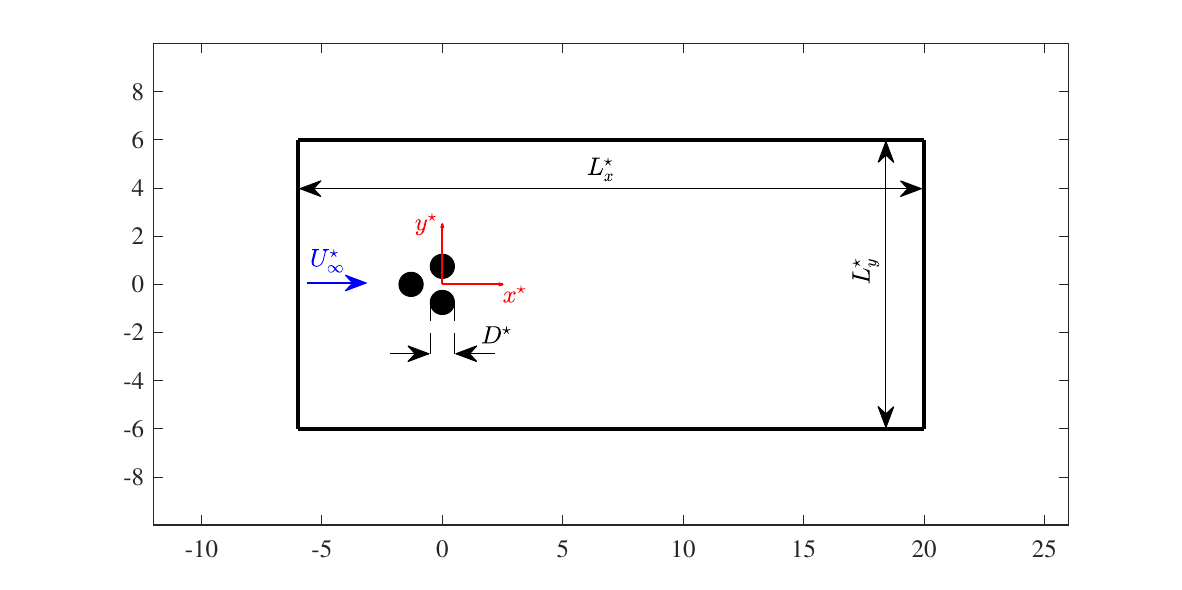}
	\caption{\label{fig:layout} Schematic representation of the fluidic pinball configuration.}
\end{figure}

The fluidic pinball is a flow configuration consisting of three rotatable cylinders of equal diameter $D^\star$, whose axes are located in the vertices of an equilateral triangle, as sketched in figure~\ref{fig:layout}. The triangle has a centre-to-centre side length equal to $1.5 D^\star$, and it is immersed in a viscous incompressible flow with a uniform upstream velocity $U^\star_\infty$. The Reynolds number for this set-up is defined as 
\begin{equation}
Re=\dfrac{\rho U^\star_\infty D^\star}{\mu},
\end{equation}
where $\rho$ and $\mu$ are the density and dynamic viscosity of the fluid, respectively. Note that all dimensional quantities, except the fluid properties ($\rho, \mu$), are denoted with the superscript $\star$.

Direct numerical simulations of the incompressible Navier–Stokes equations are used to compute the two-dimensional viscous wake behind the pinball configuration (see following \S~\ref{subsec:Navier-Stokes}). All physical quantities are made dimensionless with respect to the diameter $D^\star$ and velocity $U^\star_\infty$,
\begin{equation}
\label{eq:reference_quantities}
x=\dfrac{x^\star}{D^\star}, \quad  y=\dfrac{y^\star}{D^\star}, \quad u=\dfrac{u^\star}{U^\star_\infty}, \quad v=\dfrac{v^\star}{U^\star_\infty}, \quad p = \dfrac{2 p^\star}{\rho U^{\star 2}_\infty} \quad t=t^\star \dfrac{U^\star_\infty}{D^\star}.
\end{equation}
As shown in figure~\ref{fig:layout}, the computational domain is a rectangle excluding the interior of the cylinders, with sides equal to $L^\star_x = 26 D^\star$ and $L^\star_y = 12 D^\star$. The domain is described by a Cartesian coordinate system with origin located in the middle of the rear cylinders. The streamwise and normal-to-flow directions are denoted as $x$ and $y$, respectively. 

As rotation of the cylinders is not considered here, a no-slip boundary condition is applied on their surface. At the domain inlet (left side of the rectangle), a uniform velocity profile is prescribed, namely $u=1$ and $v=0$. A standard free-outflow boundary condition is enforced on the domain outlet (right side of the rectangle), while the remaining sides are equipped with homogeneous Neumann boundary conditions for all variables. Values of the streamwise ($u$) and normal-to-flow ($v$) velocity components prescribed as initial conditions to start the computation read as
\begin{equation}
\label{eq:init}
u(t=0)=1+u^\prime_0, \quad v(t=0)=v^\prime_0,
\end{equation}
namely as the sum of the free-stream flow velocity ($u=1$, $v=0$) and perturbation terms ($u^\prime_0$, $v^\prime_0$).
To identify the flow reduced-order coordinates from the high-dimensional simulation data through the Diffusion Maps embedding procedure described in \S~\ref{sec:methods}, the perturbations  terms $u^\prime_0$ and $v^\prime_0$ are modelled as white noise, with an amplitude equal to 10\% of the free-stream velocity.

\subsection{The incompressible Navier–Stokes equations}
\label{subsec:Navier-Stokes}

The incompressible Navier–Stokes equations are solved by means of the open-source code BASILISK (\url{http://basilisk.fr}), which implements a second-order accurate finite-volume scheme. By employing the reference dimensional quantities listed in (\ref{eq:reference_quantities}), the dimensionless form of the governing equations is obtained, reading as
\begin{subequations}
	\begin{eqnarray}
	\dfrac{\partial u}{\partial x} + 	\dfrac{\partial v}{\partial y}&=&0 \label{eq:continuity},
	\\
	\dfrac{\partial u}{\partial t} + u \dfrac{\partial u}{\partial x} + v \dfrac{\partial u}{\partial y}&=& -\dfrac{1}{\rho}\dfrac{\partial p}{\partial x} + \dfrac{1}{Re}\bigg(\dfrac{\partial^2 u}{\partial x^2} + \dfrac{\partial^2 u}{\partial y^2}\bigg) \label{eq:momentum_u},
	\\
	\dfrac{\partial v}{\partial t} + u \dfrac{\partial v}{\partial x} + v \dfrac{\partial v}{\partial y}&=& -\dfrac{1}{\rho}\dfrac{\partial p}{\partial y} + \dfrac{1}{Re}\bigg(\dfrac{\partial^2 v}{\partial x^2} + \dfrac{\partial^2 v}{\partial y^2}\bigg) \label{eq:momentum_v}.
	\end{eqnarray}
\end{subequations}
%\begin{subequations}
%	\begin{eqnarray}
%	\dfrac{\partial u}{\partial x} + 	\dfrac{\partial v}{\partial y}&=&0 \label{eq:continuity},
%	\\
%	\dfrac{\partial u}{\partial t} + u \dfrac{\partial u}{\partial x} + v \dfrac{\partial u}{\partial y}&=& -\dfrac{1}{\rho}\dfrac{\partial p}{\partial x} + \dfrac{\mu}{\rho}\bigg(\dfrac{\partial^2 u}{\partial x^2} + \dfrac{\partial^2 u}{\partial y^2}\bigg),
%	\\
%	\dfrac{\partial v}{\partial t} + u \dfrac{\partial v}{\partial x} + v \dfrac{\partial v}{\partial y}&=& -\dfrac{1}{\rho}\dfrac{\partial p}{\partial y} + \dfrac{\mu}{\rho}\bigg(\dfrac{\partial^2 v}{\partial x^2} + \dfrac{\partial^2 v}{\partial y^2}\bigg).
%	\end{eqnarray}
%\end{subequations}
%\begin{subequations}
%	\begin{eqnarray}
%	\dfrac{\partial u}{\partial t} + u \dfrac{\partial u}{\partial x} + v \dfrac{\partial u}{\partial y}&=& \dfrac{\mu}{\rho}\bigg(\dfrac{\partial^2 u}{\partial x^2} + \dfrac{\partial^2 u}{\partial y^2}\bigg),
%	\\
%	\dfrac{\partial v}{\partial t} + u \dfrac{\partial v}{\partial x} + v \dfrac{\partial v}{\partial y}&=& \dfrac{\mu}{\rho}\bigg(\dfrac{\partial^2 v}{\partial x^2} + \dfrac{\partial^2 v}{\partial y^2}\bigg).
%	\end{eqnarray}
%\end{subequations}
As usual for incompressible flows, Eqs.(\ref{eq:continuity})-(\ref{eq:momentum_v}) are solved by means of the so-called projection method. In this procedure, a temporary velocity field is first found by ignoring the pressure gradient. In a second step, the temporary field is projected onto a space of divergence-free velocity fields by adding the appropriate pressure gradient correction. For a detailed description of the numerical schemes implemented in BASILISK, the reader can refer to \cite{Popinet2003}.

All the computations are performed on a uniform structured grid with mesh size equal to $\Delta x = \Delta y = 0.2$, i. e. 5 grid cells within the length scale $D^\star$. We verified such a spatial resolution to be sufficient to achieve the grid-independence of results in all the regimes considered.

%in BASILISK by combining a second-order staggered-in-time discretization with a time-splitting projection method. This gives the following time-stepping scheme:
%\begin{eqnarray}
%\rho_{n+\frac{1}{2}}\left(\frac{\mathbf{u}_{*}-\mathbf{u}_{n}}{\Delta t}+\mathbf{u}_{n+\frac{1}{2}} \cdot \boldsymbol{\nabla} \mathbf{u}_{n+\frac{1}{2}}\right)&=&\boldsymbol{\nabla} \cdot\left(\eta_{n+\frac{1}{2}} \mathbf{D}_{*}\right)-\boldsymbol{\nabla} p_{n-\frac{1}{2}} \label{eq:2} \\
%\mathbf{u}_{n+1}&=&\mathbf{u}_{*}-\frac{\Delta t}{\rho_{n+\frac{1}{2}}}\left(\boldsymbol{\nabla} p_{n+\frac{1}{2}}-\boldsymbol{\nabla} p_{n-\frac{1}{2}}\right) \label{eq:3} \\
%\boldsymbol{\nabla} \cdot \mathbf{u}_{n+1}&=&0 \label{eq:4}
%\end{eqnarray}
%Combining~(\ref{eq:3}) and~(\ref{eq:4}) results in the following Poisson equation
%\begin{equation}
%\boldsymbol{\nabla} \cdot\left(\frac{\Delta t}{\rho_{n+\frac{1}{2}}} \nabla p_{n+\frac{1}{2}}\right)=\boldsymbol{\nabla} \cdot\left(\mathbf{u}_{*}+\frac{\Delta t}{\rho_{n+\frac{1}{2}}} \nabla p_{n-\frac{1}{2}}\right)
%\end{equation}
%The momentum equation~(\ref{eq:2}) can be reorganized as
%\begin{equation}
%\frac{\rho_{n+\frac{1}{2}}}{\Delta t} \mathbf{u}_{\star}-\boldsymbol{\nabla} \cdot\left(\eta_{n+\frac{1}{2}} \mathbf{D}_{*}\right)=\rho_{n+\frac{1}{2}}\left[\frac{\mathbf{u}_{n}}{\Delta t}-\mathbf{u}_{n+\frac{1}{2}} \cdot \nabla \mathbf{u}_{n+\frac{1}{2}}\right]-\nabla p_{n-\frac{1}{2}}
%\end{equation}

\subsection{Numerical Simulations and Flow regimes}
\label{subsec:simulation}
Spatio-temporal numerical simulations of the two-dimensional wake flow developing past the cylinders array are performed in the range $Re \in [1,130]$; an overview of the main analyzed cases is reported in Table~\ref{tab:Flow_regime}. We focus our attention on the vorticity field $\omega$, which is stored with a time-step $\Delta t=0.5$ for a total computational time equal to $T=2000$. Therefore, 4000 temporal realizations of the two-dimensional vorticity field for each simulation are considered.

\begin{table}
    \begin{center}
        \renewcommand{\arraystretch}{1.5} % Adjust the value as needed
        \begin{tabular}{lcc}
            \hline
            \textbf{Flow regime} & \textbf{Reynolds number range} & \textbf{Selected $Re$ value} \\
            \hline
            Steady symmetric & $Re < 18$ & $10$ \\
            Periodic symmetric & $18 < Re < 68$ & $30$ \\
            Periodic asymmetric & $68 < Re < 104$ & $80$ \\
            Quasi-periodic symmetric & $104 < Re < 115$ & $105$ \\
            Chaotic statistically symmetric & $Re > 115$ & $130$ \\
            \hline
        \end{tabular}
    \end{center}
    \caption{Flow regimes exhibited by the fluidic pinball as a function of the Reynolds number $Re$ with selected values investigated in the present study.}
    \label{tab:Flow_regime}
\end{table}

\begin{figure}%[h]
	\centering
	\includegraphics[scale=0.8]{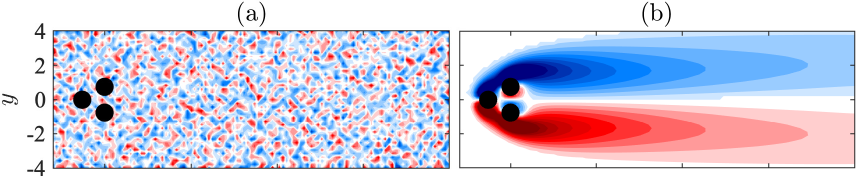}
	\includegraphics[scale=0.8]{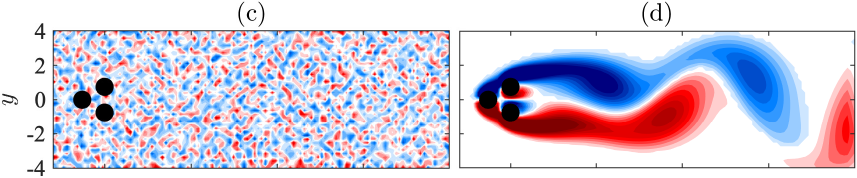}
	\includegraphics[scale=0.8]{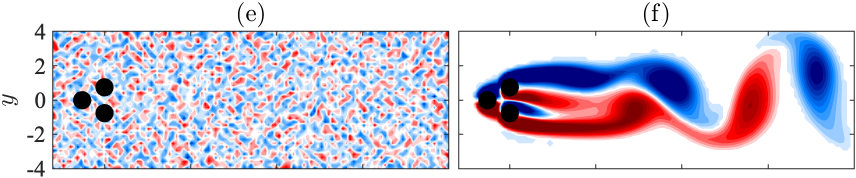}
	\includegraphics[scale=0.8]{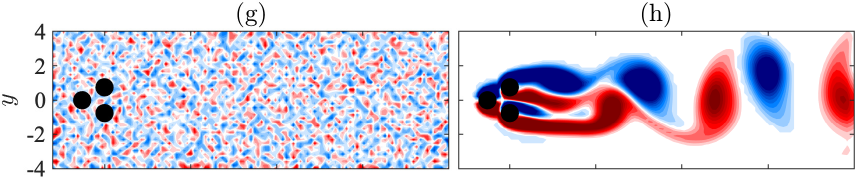}
	\includegraphics[scale=0.8]{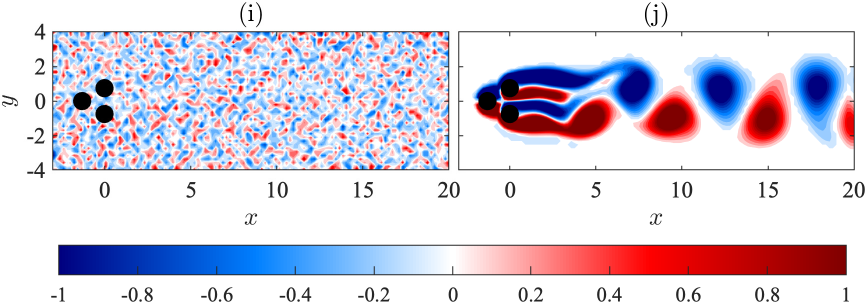}
	\caption{\label{fig:omega_contour_var_Re} Initial condition (left panels) and instantaneous snapshot after the transient (right panels) of the vorticity field $\omega(x,y)$ by varying the Reynolds number: $Re = 10$ ((a)-(b)); $30$ ((c)-(d)); $80$ ((e)-(f)); $105$ ((g)-(h)); $130$ ((i)-(j)).}
\end{figure}

\begin{figure}%[h]
	\centering
	\includegraphics[scale=0.8]{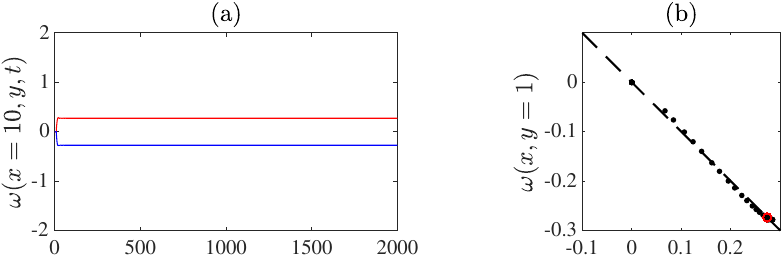}\\
	\vspace{0.1cm}
	\includegraphics[scale=0.8]{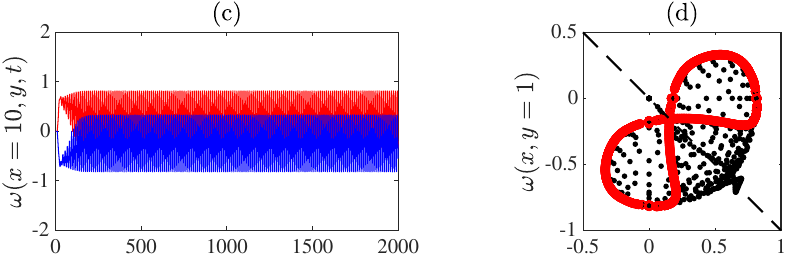}\\
	\vspace{0.1cm}
	\includegraphics[scale=0.8]{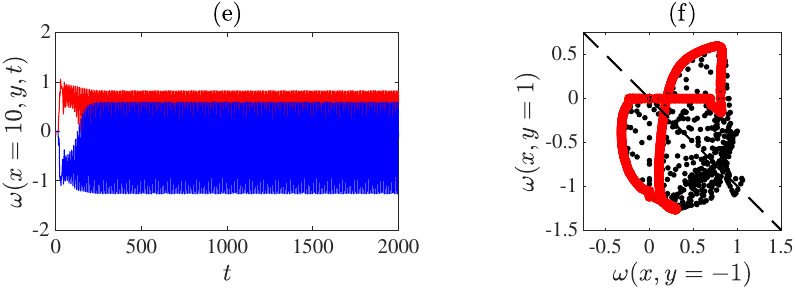}\\
	\vspace{0.1cm}
	\includegraphics[scale=0.8]{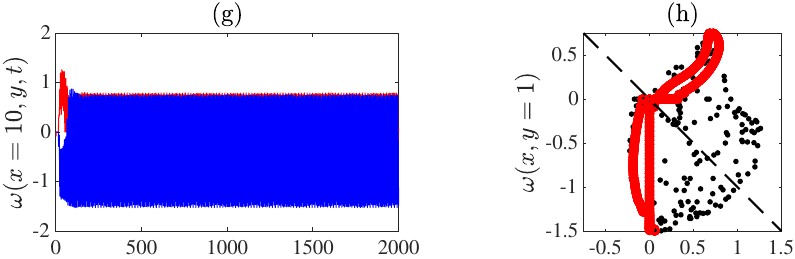}\\
	\vspace{0.1cm}
	\includegraphics[scale=0.8]{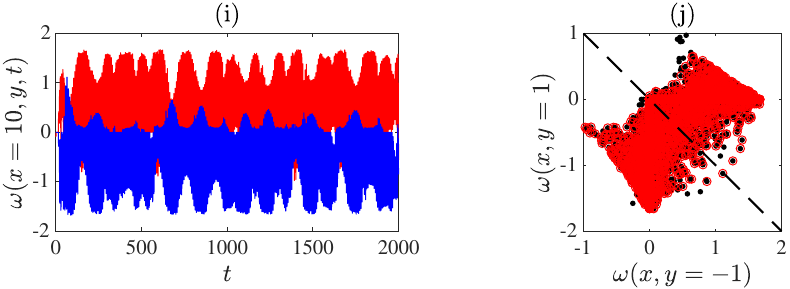}
	\caption{\label{fig:omega_FFT_var_Re} Left panels: temporal evolution of vorticity $\omega(x = 10, y, t)$ at the vertical locations $y=-1$ (red curves) and $y=1$ (blue curves). Right panels: $\omega(x = 10, y=1, t)$ plotted as a function of $\omega(x = 10, y=-1, t)$ including (black dots) and excluding (red circles) the transient, reporting also the -1 slope line (black dashed curve). From top to bottom: $Re = 10$ ((a)-(b)); $30$ ((c)-(d)); $80$ ((e)-(f)); $105$ ((g)-(h)); $130$ ((i)-(j)).
	}
\end{figure}

Snapshots of numerical simulations are shown in figures~\ref{fig:omega_contour_var_Re}-\ref{fig:omega_FFT_var_Re} in terms of instantaneous two-dimensional contour maps (figure~\ref{fig:omega_contour_var_Re}) and time-resolved signals (figure~\ref{fig:omega_FFT_var_Re}) of the vorticity field. 
In particular, figure~\ref{fig:omega_contour_var_Re} shows the initial condition (left panels) and the instantaneous vorticity field after the transient (right panels) over the entire spatial domain, by increasing the Reynolds number from $Re=10$ ((a)-(b)) to $Re=130$ ((i)-(j)). 
The left panels of figure~\ref{fig:omega_FFT_var_Re} illustrate the temporal evolution of the vorticity field at two symmetric vertical locations in the flow, namely $y=-1$ and $1$, at the streamwise station $x=10$. The two vorticity signals are then plotted one versus the other in the right panels of figure~\ref{fig:omega_FFT_var_Re}. Again, the Reynolds number increases from $Re=10$ ((a)-(b)) to $Re=130$ ((i)-(j)). 

After a fast transient dynamics depending on the random perturbations employed to initialize the computations, we retrieve in the present results the five different flow regimes described by \cite{Deng_Noack_2020}. For $Re=10$, a steady flow field is obtained after a computational time $t\approx 20 \Delta t$ (see figure~\ref{fig:omega_contour_var_Re}(b)). The vorticity signals reported in figure~\ref{fig:omega_FFT_var_Re}(a)-(b) show a quick convergence of the dynamics excited by the initial condition to a ``fixed-point'' mirror-symmetric solution with respect to the $y=0$ axis (red circle in (b)), which corresponds to $\omega(x=10,y=-1) = -\omega(x=10,y=1) \approx 0.27$.

By increasing the Reynolds number up to $Re=30$, the flow experiences a transition from a steady to a periodic symmetric vortex shedding regime (see figure~\ref{fig:omega_contour_var_Re}(d) and figure~\ref{fig:omega_FFT_var_Re}(c)-(d)). At $Re=30$, the reduced frequency of the shedding $f = f^\star D^\star/U^\star_\infty$, where $f^\star$ is the peak frequency of the Fast Fourier Transform (FFT) of signals reported in figure~\ref{fig:omega_FFT_var_Re}(c), is equal to $f=0.085$. After a transient of approximately $t\approx 150 \Delta t$, the flow dynamics converges to a symmetric ``limit-cycle'' solution (red circles in figure~\ref{fig:omega_FFT_var_Re}(d)). The limit-cycle curve is symmetric with respect to the -1 slope line (black dashed line), thus outlining a ``spatio-temporal'' symmetry of the flow field: the absolute value of the vorticity signal at $(x=10, y=1)$ is exactly retrieved at $(x=10, y=-1)$ after a constant time delay, which depends on the oscillation period of the vortex shedding behind the cylinders.

At $Re=80$, the dynamics in characterized by an asymmetric vortex shedding with a base-bleeding jet in the near wake of the cylinders, which is reached after a transient of $t\approx 200 \Delta t$. In the present simulation, the base-bleeding jet appears deflected downward (see figure~\ref{fig:omega_contour_var_Re}(f)). The reduced frequency of the shedding slightly increases with respect to the previous regime, and it is equal to $f=0.105$. The lost of spatio-temporal symmetry in the flow dynamics is highlighted by the signals reported in figure~\ref{fig:omega_FFT_var_Re}(e)-(f), where it can be clearly seen that the two lobes of the limit-cycle solution do no more perfectly mirror on each other with respect to the -1 slope line.

As the Reynolds number reaches $Re=105$, the flow enters a quasiperiodic asymmetric regime after a transient of $t\approx 200 \Delta t$ (see figure~\ref{fig:omega_contour_var_Re}(h)). In this regime, the dynamics is characterized by a ``frequency-locking" phenomenon: along with the reduced frequency of the ``far-field" shedding, which is equal to $f=0.121$, a one-order-of-magnitude lower frequency ($f=0.01$) appears in the FFT spectra of the vorticity field (not shown), which is related to the base-bleeding jet oscillations around its deflected position. The time series reported in figure~\ref{fig:omega_FFT_var_Re}(g)-(h) display again a asymmetric solution, and a ``shrinking" effect of the two lobes of the limit cycle is observed.
 
At $Re =130$, the flow reaches a fully statistically-symmetric chaotic regime, with the base-bleeding jet oscillating randomly between upward and downward deflections (see figure~\ref{fig:omega_contour_var_Re}(j)). Due to the multi-scale spatio-temporal behaviour of the flow in this conditions, the peak frequency of the FFT spectra of the vorticity field (not shown) becomes broader and more difficult to detect; it is approximately equal to $f=0.147$. The temporal signals reported in figure~\ref{fig:omega_FFT_var_Re}(i)-(j) show the inherently high-dimensional nature of the chaotic dynamics: a limit-cycle solution can no more be detected, while a high-dimensional attractor fully covers the plane (red circles in figure~\ref{fig:omega_FFT_var_Re}(j)). 
Note that, according to the statistically-symmetric nature of chaos in this flow regime, the high-dimensional attractor mirrors with respect of the -1 slope line.
 
 \begin{figure}%[h]
 	\centering
 	\includegraphics[scale=0.8]{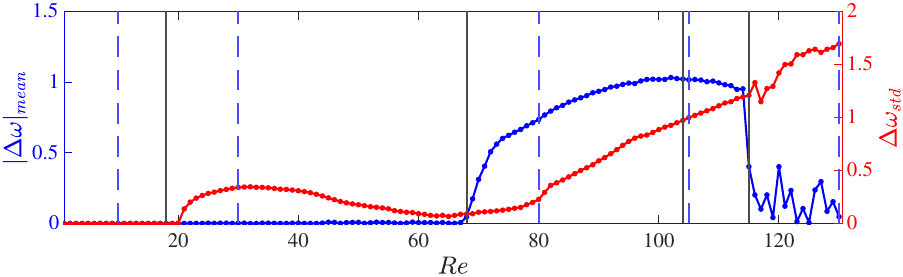}\\
 	\caption{\label{fig:Bifurcation_diagram} Mean absolute value (blue curve) and standard deviation (red curve) of the quantity $\Delta \omega$ (see~(\ref{eq:Deltaom})) by varying the Reynolds number $Re$. The cases $Re=10$, $30$, $80$, $105$ and $130$ discussed so far are highlighted by vertical blue dashed lines, while the vertical black continuos lines denote the regime transition thresholds $Re \approx 18$ (from steady symmetric to periodic symmetric), $Re \approx 68$ (from periodic symmetric to periodic asymmetric), $Re \approx 104$ (from periodic to quasi-periodic asymmetric) and $Re \approx 115$ (from quasi-periodic asymmetric to chaotic statistically symmetric) reported by \cite{Deng_Noack_2020}. 
 	}
 \end{figure}

%span all flow regimes outlined by Deng \textit{et al.}~\cite{Deng_Noack_2020}, from steady symmetric ($Re < 18$) to chaotic ($Re > 115$) behaviour
An overview of the flow transitions from the steady symmetric up to the chaotic regime is shown in figure~\ref{fig:Bifurcation_diagram}, which reports the mean and standard deviation of the quantity $\Delta \omega$ in the whole range $Re \in [1,130]$ with resolution $\Delta Re = 1$. Note that data reported in figure~\ref{fig:Bifurcation_diagram} are computed at each $Re$ by initializing the flow with the post-transient (i.e. asymptotic) solution obtained at the previous $Re$. At each time instant, $\Delta \omega$  is defined as
\begin{equation}
\label{eq:Deltaom}
\Delta \omega (t)= |\omega(\bar{x},\bar{y},t)| - |\omega(\bar{x},-\bar{y},t)|,
\end{equation}
thus quantifying the difference between the absolute value of the vorticity signals stored at two symmetric spatial locations in the flow. In particular, we selected two points located in the near-field region, i.e. $\bar{x} = 2.75$ and $\bar{y} = 1$. Therefore, the mean value $|\Delta \omega|_{mean}$ and the standard deviation $\Delta \omega_{std}$ represent a measure of the symmetry and the unsteadiness of the flow, respectively, by variation of the Reynolds number $Re$. Note that the transition thresholds between the different flow regimes, namely $Re \approx 18$, $68$, $104$ and $115$ (black continuous vertical lines in figure~\ref{fig:Bifurcation_diagram}), are the same as those reported in \cite{Deng_Noack_2020}.

\section{Parsimonious Diffusion Maps embedding}
\label{sec:methods}

Here, we describe, the nonlinear manifold learning methodology employed to find the underlying manifold on which the fluidic pinball emergent dynamics evolves by variation of the Reynolds number $Re$, namely parsimonious diffusion maps (\cite{dsilva2018parsimonious,galaris2022numerical,Patsatzis_2023,gallos2024data}. The goal is to construct a non-linear mapping from a high-dimensional space to a  parsimonious low-dimensional subspace, so that the intrinsic geometry of the embedded manifold is preserved. We employ the theoretical derivation and numerical implementation described in \cite{dsilva2018parsimonious,holiday2019manifold,Patsatzis_2023,gallos2024data}.

Let $\boldsymbol{\omega}_m \in \mathbb{R}^N$, $m = 1, \ldots, M$ be the collected $M$ observations in the high-dimensional space, which represent the different temporal realizations (snapshots) of the vorticity field in the whole spatial domain. The collection of snapshots can be compactly written in a row-wise matrix form, reading as $\boldsymbol{\Omega} \in \mathbb{R}^{M \times N}$.  Let us now assume that these data are embedded in a smooth low-dimensional manifold $\mathcal{M} \subset \mathbb{R}^N$. Diffusion maps seek to find their low-dimensional representations $\boldsymbol{\psi}_m \in \mathbb{R}^D$ with $D \ll N$, compactly written in the matrix form $\boldsymbol{\Psi} \in \mathbb{R}^{M \times D}$, such that the Euclidean distance of the points $\boldsymbol{\psi}$ (rows of $\boldsymbol{\Psi}$) is preserved as the diffusion distance of points in $\boldsymbol{\omega}$ (rows of $\boldsymbol{\Omega}$) (\cite{nadler2006diffusion}).

The implementation begins by defining a similarity metric between pairs of points, say $\boldsymbol{\omega}_i, \boldsymbol{\omega}_j \in \boldsymbol{\Omega}$, $\forall i, j = 1, \ldots, M$ in the high-dimensional space. By utilizing the Euclidean norm $d_{ij} = \|\boldsymbol{\omega}_i - \boldsymbol{\omega}_j\|$, a Gaussian kernel function $k(\boldsymbol{\omega}_i, \boldsymbol{\omega}_j)$ is employed for calculating the affinity matrix, which reads as
\begin{equation}
\textbf{A} = [a_{ij}] = [k(\boldsymbol{\omega}_i, \boldsymbol{\omega}_j)] = \exp\left(-\frac{{d_{ij}^2}}{\epsilon^2}\right) = \exp\left(-\frac{{\|\boldsymbol{\omega}_i - \boldsymbol{\omega}_j\|^2}}{\epsilon^2}\right),
\label{eq:affinity_matrix}
\end{equation}
where the shape parameter $\epsilon$ expresses a measure of the local neighbourhood in the high-dimensional space.

Then, one constructs the $M \times M$ Markovian transition matrix $\textbf{M}$, by normalizing each row of the affinity matrix, such that
\begin{equation}
\mathbf{M}=\mathbf{D}^{-1} \mathbf{A}, \quad \text { where } \quad \mathbf{D}=\operatorname{diag}\left(\sum_{j=1}^{M} a_{i j}\right).
\label{eq:Markovian_matrix}
\end{equation}
The elements $\mu_{i j}$ of $\mathbf{M}$ correspond to the probability of jumping from one point to another in the high-dimensional space. In  particular, the transition matrix defines a Markovian random walk $\Omega_{t}$ on the data points, with probability of moving from  point $i$ to point $j$ at the $t$-step of a conceptual diffusion process across the data,
\begin{equation}
\mu_{i j}=\mu\left(\boldsymbol{\omega}_{i}, \boldsymbol{\omega}_{j}\right)=\operatorname{Prob}\left(\Omega_{t+1}=\boldsymbol{\omega}_{j} \mid \Omega_{t}=\boldsymbol{\omega}_{i}\right).
\label{eq:transtion}
\end{equation}
Given the weighted graph defined via the Gaussian kernel function $k\left(\boldsymbol{\omega}_{i}, \boldsymbol{\omega}_{j}\right)$ on the high-dimensional space, the random  walk can be then defined by the transition probabilities
\begin{equation}
\mu_{i j}=\mu\left(\boldsymbol{\omega}_{i}, \boldsymbol{\omega}_{j}\right)=\frac{k\left(\boldsymbol{\omega}_{i}, \boldsymbol{\omega}_{j}\right)}{\operatorname{deg}\left(\boldsymbol{\omega}_{i}\right)}, \quad \text { where } \operatorname{deg}\left(\boldsymbol{\omega}_{i}\right)=\sum_{j=1}^{M} k\left(\boldsymbol{\omega}_{i}, \boldsymbol{\omega}_{j}\right),
\end{equation}
thus retrieving the expression given in~(\ref{eq:Markovian_matrix}).

The Markovian transition matrix $\mathbf{M}$ is similar to the matrix $\hat{\mathbf{M}}=\mathbf{D}^{-1 / 2} \mathbf{A} \mathbf{D}^{-1 / 2}$, which is symmetric and positive definite,  thus allowing for the eigendecomposition
\begin{equation}
\mathbf{M}=\sum_{i=1}^{M} \lambda_{i} \mathbf{w}_{i} \mathbf{u}_{i}^\top,
\end{equation}
where $\lambda_{i} \in \mathbb{R}$ are the eigenvalues and $\mathbf{w}_{i} \in \mathbb{R}^{M}$ and $\mathbf{u}_{i} \in \mathbb{R}^{M}$ are the left and right eigenvectors of $\mathbf{M}$, respectively, such that $\left\langle\mathbf{w}_{i}, \mathbf{u}_{j}\right\rangle=\delta_{i}^{j}$. The set of right eigenvectors $\mathbf{u}_{i}$ provides an orthonormal basis for the low-dimensional subspace in $\mathbb{R}^{D}$ spanned by the rows of $\mathbf{M}$. The best $D$-dimensional low-rank approximation of the row space of $\mathbf{M}$ in the Euclidean space  $\mathbb{R}^{D}$ is given by the $D$ right eigenvectors corresponding to the $D$ largest eigenvalues. 
%\textcolor{red}{DP: Since the embedding is eigenvalues times eigenvectors, I kept $\psi$ as the embedding to avoid change of the figures later on and changed the evecs symbols.}

Thus, the standard DMs embedding is realized as the mapping of each observation $\boldsymbol{\omega}_{m}$ to the row vector
\begin{equation}
\boldsymbol{\psi}_{m}=\left(\lambda_1 u_{1, m}, \ldots, \lambda_D u_{D, m}\right), \forall m=1, \ldots, M,
\end{equation}
where $u_{i, m}$ denotes the $m$-th element of the $i$-th right eigenvector, corresponding to the $i$-th non-trivial, sorted in descending order, eigenvalue $\lambda_{i}$ for $i=1, \ldots, D \ll N$. It has been shown that such a selection of DMs embeddings provides the best  approximation of the Euclidean distance $\left\|\boldsymbol{\psi}_{i}-\boldsymbol{\psi}_{j}\right\|$ between two points, say $\boldsymbol{\psi}_{i}, \boldsymbol{\psi}_{j} \in \boldsymbol{\Psi}$, on the low-dimensional space, to the  diffusion distance on the high-dimensional space (\cite{nadler2006diffusion}), which is defined as
\begin{equation}
D_{t}^{2}\left(\boldsymbol{\omega}_{i}, \boldsymbol{\omega}_{j}\right)=\left\|\mu_{t}\left(\boldsymbol{\omega}_{i}, \cdot\right), \mu_{t}\left(\boldsymbol{\omega}_{j}, \cdot\right)\right\|_{L_{2}, 1 / \operatorname{deg}}^{2}=\sum_{k=1}^{M} \frac{\left(\mu_{t}\left(\boldsymbol{\omega}_{i}, \boldsymbol{\omega}_{k}\right)-\mu_{t}\left(\boldsymbol{\omega}_{j}, \boldsymbol{\omega}_{k}\right)\right)^{2}}{\operatorname{deg}\left(\boldsymbol{\omega}_{k}\right)},
\end{equation}
where $\mu_{t}\left(\boldsymbol{\omega}_{i}, \cdot\right)$ is the $i$-th row of the Markovian transition matrix $\mathbf{M}^{t}$, corresponding to the transition probabilities after $t$ diffusion steps. In our computations, we considered $t=1$.

In practice, the embedded dimension $D$ is determined by the spectral gap of the eigenvalue ratio of the transition matrix  $\mathbf{M}$, assuming that the first $D$ leading eigenvalues are adequate for providing a good approximation of the diffusion distance  between all pairs of points (\cite{coifman2008diffusion}). However, this is not always the case, since some of the first $D$ eigenvectors may be higher harmonics of previous ones and thus they do not describe new directions along the data set. 

To consider these cases, we have further employed parsimonious DMs \cite{dsilva2018parsimonious,holiday2019manifold,Patsatzis_2023,gallos2024data} to select the eigenvectors that provide unique directions along the data set, thus providing the best $D$-dimensional embedding. Given the set $\mathbf{u}_1,\ldots,\mathbf{u}_{k-1}$ of the first $k-1$ DM eigenvectors, we use a local linear regression model to fit the $k$-th eigenvector $\mathbf{u}_k$ against all the previous ones, for each element $m=1,\ldots,M$ as
\begin{equation}
    u_{k,m} \approx \alpha_{k,m} + \boldsymbol{\beta}_{k,m}^\top \mathbf{U}_{k-1,m}, 
\end{equation}
where $\alpha_{k,m}\in\mathbb{R}$, $\boldsymbol{\beta}_{k,m}\in \mathbb{R}^{k-1}$ and $\mathbf{U}_{k-1,m}=[u_{1,m}, \ldots, u_{k-1,m}]^\top$. The parameters $\alpha_{k,m}$ and $\boldsymbol{\beta}_{k,m}$ are found from the solution of the following optimization problem:
\begin{equation}
    (\alpha_{k,m}, \boldsymbol{\beta}_{k,m}) = \underset{\alpha,\boldsymbol{\beta}}{argmin} \sum_{i\neq m}exp\left(-\dfrac{{\|\mathbf{U}_{k-1,m} - \mathbf{U}_{k-1,i}\|^2}}{\epsilon^2} \right) \left(u_{k,i}-(\alpha + \boldsymbol{\beta}^\top \mathbf{U}_{k-1,i})\right).
\end{equation}
Then, the normalized leave-out-one-cross-validation error is measured by the local linear fitting coefficient as
\begin{equation}
    r_k = \sqrt{\dfrac{\sum_{m=1}^M \left(u_{k,m}-(\alpha_{k,m} + \boldsymbol{\beta}_{k,m}^\top \mathbf{U}_{k-1,m})\right)^2}{\sum_{m=1}^M u_{k,m}^2}}.
    \label{eq:LLFc}
\end{equation}
With the definition~(\ref{eq:LLFc}), a small $r_k$ indicates that the $k$-th eigenvector $\mathbf{u}_k$ does not provide a new direction along the data set, since it is well approximated by the first $k-1$ DM eigenvectors $\mathbf{u}_1,\ldots,\mathbf{u}_{k-1}$. In other terms, including $\mathbf{u}_k$ to the embedding would not significantly improve the parametrization of the manifold. Hence, we select the DMs embeddings with the highest $r_k$ values.

%However, this is not always the case and one should consider employing parsimonious DMs  (\cite{dsilva2018parsimonious}, \cite{holiday2019manifold}) for selecting the eigenvectors that provide the best low-dimensional embedding. For the cases considered in this work (see following \S~\ref{sec:results}), though, such a parsimonious identification was not required, since the spectral gap was an adequate criterion for discovering the DMs eigenvectors that provide the desired manifold parametrization.

\subsection{The Encoder for out-of-sample extensions}
\label{subsec:restriction}
Let the high-dimensional data set be $\boldsymbol{\Omega}=\left\{\boldsymbol{\omega}_{m} \in \mathbb{R}^{N} \mid m=1, \ldots, M\right\}$ of $M$ observations and assume that the employment of DMs results in a low-dimensional embedding with, say, $\lambda_{i}$ for $i=1, \ldots, D$ sorted eigenvalues and their  associated right eigenvectors $\boldsymbol{\psi}_{i}$. Here, the restriction operator $\mathcal{R}$, evaluated at a point of the given data set $\boldsymbol{\omega}_{m}$, is defined as
\begin{equation}
\mathcal{R}\left(\boldsymbol{\omega}_{m}\right)=\left(\lambda_1 u_{1, m}, \ldots, \lambda_D u_{D, m}\right)=\boldsymbol{\psi}_{m} \in \mathbb{R}^{D}, \quad m=1, \ldots M,
\label{eq:restriction}
\end{equation}
where $u_{i, m}$ denotes the $m$-th component of the $i$-th DMs eigenvector $\mathbf{u}_{i}$. However, in the case where new unseen high-dimensional data points are presented, one needs to recalculate the DMs embeddings. To avoid the computational cost of such a procedure, we instead utilize the so-called Nyström method (\cite{nystrom1929uber}) to extend $\mathcal{R}$ to new unseen data points (\cite{coifman2008diffusion}), i.e. solve the out-of-sample extension problem. As we will discuss in \S~\ref{subsec:recon_err}, this represents a key point for the evaluation of the error between the ``ground truth'' solution, namely the high-fidelity simulation data, and the ``reconstructed'' values of the DMs procedure.

The extension problem is based on the fact that the eigenvectors $\mathbf{u}_{i}$ form a basis of the low-dimensional subspace (see \cite{coifman2006geometric} for details). This implies that any function defined at known data points $\boldsymbol{\omega}_{m}$ of the high-dimensional space, say $\boldsymbol{f}\left(\boldsymbol{\omega}_{m}\right)$, can be extended to the low-dimensional manifold as
\begin{equation}
\hat{\boldsymbol{f}}\left(\boldsymbol{\omega}_{m}\right)=\sum_{i=1}^{D} a_{i} \mathbf{u}_{i}\left(\boldsymbol{\omega}_{m}\right) \quad m=1, \ldots, M,
\label{eq:f}
\end{equation}
where $a_{i}=\left\langle\mathbf{u}_{i}, \boldsymbol{f}\left(\boldsymbol{\omega}_{m}\right)\right\rangle$ are the projection coefficients of the function to be extended to the first $D$ eigenvectors. Using the same projection coefficients, one can now map the function evaluated to new unseen data points, say $\boldsymbol{\omega}_{l}^{n} \in \mathbb{R}^{N}$ for $l=1, \ldots, L$, of the high-dimensional space as
\begin{equation}
\mathcal{E}\left(\hat{\boldsymbol{f}}\left(\boldsymbol{\omega}_{l}^{n}\right)\right)=\sum_{i=1}^{D} a_{i} \hat{\mathbf{u}}_{i}\left(\boldsymbol{\omega}_{l}^{n}\right),
\label{eq:E}
\end{equation}
where
\begin{equation}
\hat{\mathbf{u}}_{i}\left(\boldsymbol{\omega}_{l}^{n}\right)=\frac{1}{\lambda_{i}} \sum_{m=1}^{M} k\left(\boldsymbol{\omega}_{m}, \boldsymbol{\omega}_{l}^{n}\right) \mathbf{u}_{i}\left(\boldsymbol{\omega}_{m}\right), \quad i=1, \ldots, D ,
\label{eq:GHs}
\end{equation}
are the corresponding Geometric Harmonics (see following \S~\ref{subsec:lifting}), and $k\left(\boldsymbol{\omega}_{m}, \boldsymbol{\omega}_{l}^{n}\right)$ is the Gaussian kernel function used for the construction of the affinity matrix in~(\ref{eq:affinity_matrix}), this time measuring the similarity between the new point $\boldsymbol{\omega}_{l}^{n}$ and the known data points $\boldsymbol{\omega}_{m}$. Casting the data set of new points $\boldsymbol{\omega}_{l}^{n}$ for $l=1, \ldots, L$ in the matrix form $\boldsymbol{\Omega}^{n} \in \mathbb{R}^{L \times N}$, (\ref{eq:E}) can be written in compact matrix form as
\begin{equation}
\mathcal{E}(\hat{\boldsymbol{f}})=\mathbf{K}_{L \times M} \mathbf{U}_{M \times D} \boldsymbol{\Lambda}_{D \times D}^{-1} \mathbf{U}_{D \times M}^\top \boldsymbol{f}_{M \times 1},
\label{eq:E2}
\end{equation}
where $\mathbf{K}_{L \times M}$ is the corresponding kernel matrix, $\mathbf{U}_{M \times D}$ is the matrix of the DMs eigenvectors $\mathbf{u}_{i}$ in columns and $\boldsymbol{\Lambda}_{D \times D}$ is the diagonal matrix with elements $\lambda_{i}$.

\begin{figure}%[h]
	\centering
	\includegraphics[scale=0.28]{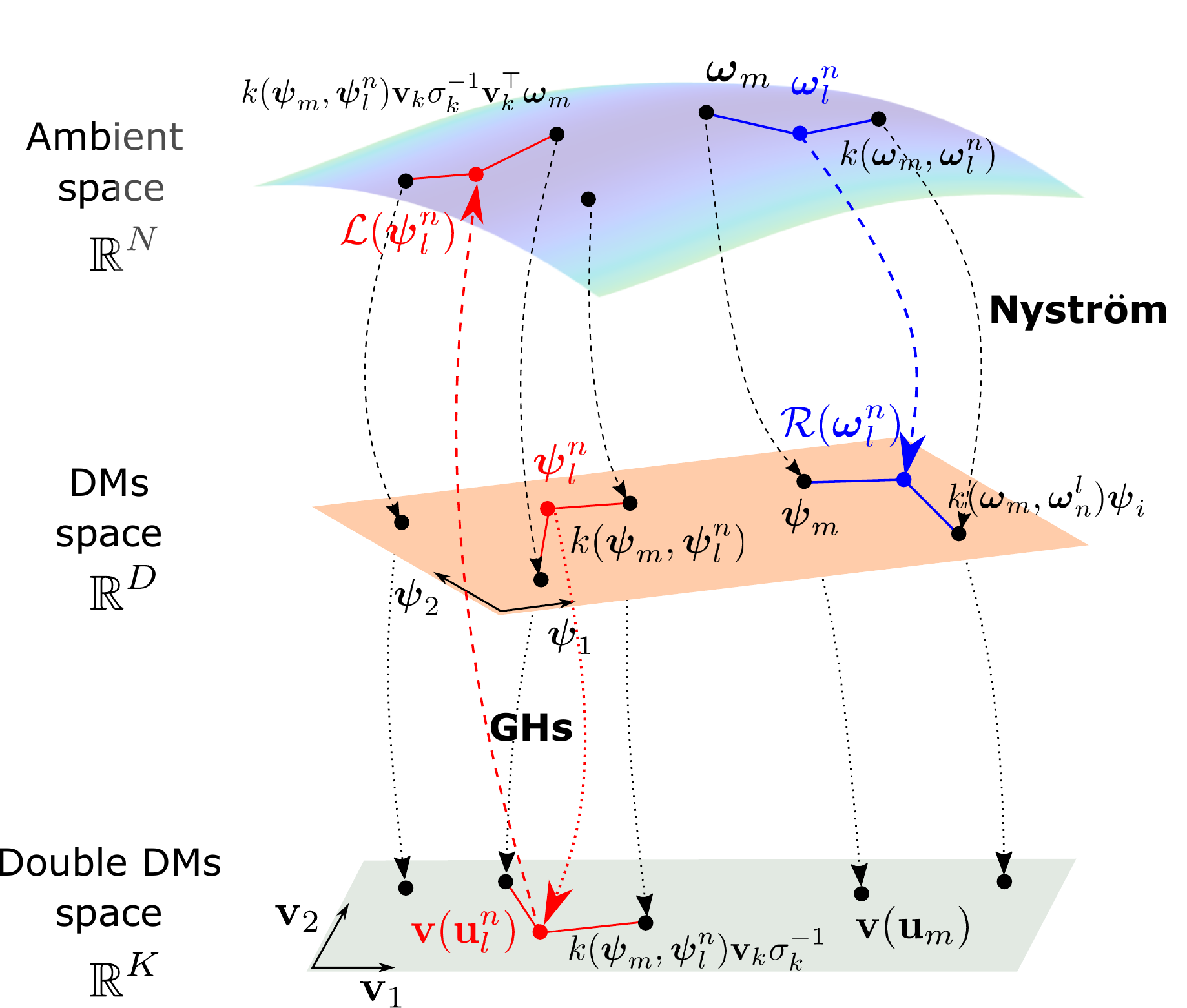}
	\caption{\label{fig:Diffusion_Maps_meth}Schematic representation of the encoding (blue dots and arrows) via the Nyström method to the DMs and the decoding operations via Geometric Harmonics and ``double`` DMs (red dots and arrows). The black dots denote observations in the ambient space $\mathbb{R}^{N}$ and their representations on the DMs and ``double" DMs spaces $\mathbb{R}^{D}$ and $\mathbb{R}^{K}$, respectively. The Nyström method is performed for obtaining the image $\mathcal{R}\left(\boldsymbol{\omega}_{l}^{n}\right)$ of the new point $\boldsymbol{\omega}_{l}^{n}$. The reconstruction with GHs is performed for obtaining the pre-image $\mathcal{L}\left(\boldsymbol{\psi}_{l}^{n}\right)$ of a new point $\boldsymbol{\psi}_{l}^{n}=\mathbf{u}_l^n \boldsymbol{\Lambda}$. As described in \S~\ref{subsec:lifting} for the GHs extension, projection to the ``double" DMs space is first performed for defining a new basis in $\mathbb{R}^{K}$ through which one obtains the reconstructed state in $\mathbb{R}^{N}$.
	}
\end{figure}

For the solution of the out-of-sample extension problem, we seek to extend the restriction operator of (\ref{eq:restriction}) for new out-of-sample unseen data points $\boldsymbol{\Omega}^{n} \not \subset \boldsymbol{\Omega}$ of the high-dimensional space. Thus, substitution of $\hat{\boldsymbol{f}}$ with $\mathcal{R}$ and of $\boldsymbol{f}$ with $\boldsymbol{\Psi}=\mathbf{U} \boldsymbol{\Lambda}$ into~(\ref{eq:E2}) (the left and right hand sides in~(\ref{eq:restriction})) implies
\begin{equation}
\label{eq_Nystrom}
\mathcal{R}\left(\boldsymbol{\Omega}^{n}\right)=\boldsymbol{K}_{L \times M} \mathbf{U}_{M \times D}.
\end{equation}
The above extends the definition of the encoder to the $L$ new data points in $\boldsymbol{\Omega}^{n}$ on the manifold as constructed by the DMs embedding. A schematic representation of the Nyström method is displayed in figure~\ref{fig:Diffusion_Maps_meth}, where the image $\mathcal{R}\left(\boldsymbol{\omega}_{l}^{n}\right)$ of the new point $\boldsymbol{\omega}_{l}^{n}$ is depicted with blue dots and arrows.

\subsection{The Decoder}
\label{subsec:lifting}

The numerical solution of the pre-image problem refers to the reconstruction of high-dimensional representations, given low-dimensional measurements on the manifold. This task corresponds to the construction of the so-called lifting operator (the decoder)
\begin{equation}
\mathcal{L} \equiv \mathcal{R}^{-1}: \mathcal{R}(\boldsymbol{\omega}) \mapsto \boldsymbol{\omega},
\label{eq:lifting}
\end{equation}
so that new (unseen) point $ \boldsymbol{\psi}_{l}^{n}\notin\left\{\boldsymbol{\psi}_{m}=\mathcal{R}\left(\boldsymbol{\omega}_{m}\right) \mid m=1, \ldots, M\right\}$ for $l=1, \ldots, L$  in the low-dimensional manifold are mapped back to the original high-dimensional space. Below, we present the 
methodology employed in this work to solve the pre-image problem, namely the Geometric Harmonics. For a detailed review and comparison of various methods, the reader is refereed to \cite{chiavazzo2014reduced}, \cite{papaioannou2022time}, \cite{evangelou2022double} and \cite{Patsatzis_2023}.

Geometric Harmonics (GHs) are sets of functions that allow the extension of a function defined on a high-dimensional space to new (unseen) points on the low-dimensional manifold (\cite{coifman2006geometric}). Their derivation is based on the Nyström method and they constitute the functions previously shown in~(\ref{eq:GHs}). Thus, one would employ~(\ref{eq:GHs}) for the new points in the manifold $\boldsymbol{\psi}_{l}^{n}$ for $l=1, \ldots, L$, with the Gaussian kernel function now considered on the low-dimensional space as $k\left(\boldsymbol{\psi}_{m}, \boldsymbol{\psi}_{l}^{n}\right)$, where $\boldsymbol{\psi}_{m} \in \mathcal{R}(\boldsymbol{\Omega})$ is the DMs embedding of the $m$-th observation. However, as pointed out in \cite{chiavazzo2014reduced} and \cite{evangelou2022double}, the basis provided by the DMs embeddings $\left(\lambda_{i}, \mathbf{u}_{i}\right)$ for $i=1, \ldots, D$ is unable to provide an accurate approximation of the function on the manifold, that is the function in~(\ref{eq:f}), which in this case takes the form $\hat{\boldsymbol{f}}\left(\boldsymbol{\psi}_{m}\right)$. Instead, one should again employ DMs, this time on the $D$ first $\mathbf{u}_{i}$ eigenvectors, in order to obtain a new basis upon which to extend $\hat{\boldsymbol{f}}$ (\cite{chiavazzo2014reduced}, \cite{evangelou2022double}). In what follows, we present the implementation of the ``Double DMs" technique for calculating the GHs functions.

Recall that the employment of the DMs to the high-dimensional data of $M$ observations over the parameter space $\boldsymbol{\Omega}=\left\{\boldsymbol{\omega}_{m} \in \mathbb{R}^{N} \mid m=1, \ldots, M\right\}$ results into the embedding $\left(\lambda_{i},\mathbf{u}_{i}\right)$ for $i=1, \ldots, D$. At the first step, similarly to the application of the DMs, a Gaussian kernel function is utilized for constructing the affinity matrix for the $\mathbf{u}_{i}$ DMs eigenvectors
\begin{equation}
\tilde{\mathbf{A}}=\left[\tilde{a}_{i j}\right]=\left[k\left(\mathbf{u}_{i}, \mathbf{u}_{j}\right)\right]=\exp \left(-\frac{\left\|\mathbf{u}_{i}-\mathbf{u}_{j}\right\|^{2}}{\tilde{\epsilon}^{2}}\right),
\end{equation}
where $\tilde{\epsilon} \ll \epsilon$ of the ``first" DMs embedding. Here, one does not need to calculate a Markovian transition matrix as in the ``first" DMs, since the accurate approximation of the diffusion distance is no longer needed. Since $\tilde{\mathbf{A}}$ is a positive and semidefinite matrix, it has a set of non-negative eigenvalues, say $\sigma_{k} \in \mathbb{R}$, and the corresponding orthonormal right eigenvectors $\mathbf{v}_{k} \in \mathbb{R}^{M}$. The selection of the first $K$ largest eigenvalues, such that $\sigma_{K}>\delta \sigma_{0}$, where $\delta>0$ serves as a threshold, determines the accuracy of the new embedding; the smaller/largest $\delta$ is, the more/less accurate the ``second" DMs embedding is. We then use the eigenvectors $\mathbf{v}_{k}$ as a basis to project the function $\hat{\boldsymbol{f}}$ via the new low-dimensional points $\boldsymbol{\psi}_{l}^{n} \in \mathbb{R}^{D}$ for $l=$ $1, \ldots, L$. Since the ``first'' DMs embedding implies $\boldsymbol{\psi}_{l}^{n}=\mathbf{u}^n_l \boldsymbol{\Lambda}$, the projection of~(\ref{eq:E}) now reads
\begin{equation}
\mathcal{E}\left(\hat{\boldsymbol{f}}\left(\boldsymbol{\psi}_{l}^{n}\right)\right)=\sum_{k=0}^{K} \tilde{a}_{k} \hat{\mathbf{v}}_{k}\left(\mathbf{u}_{l}^{n}\right),
\end{equation}
where
\begin{equation}
\hat{\mathbf{v}}_{k}\left(\mathbf{u}_{l}^{n}\right)=\frac{1}{\sigma_{k}} \sum_{m=1}^{M} k\left(\boldsymbol{\psi}_{m}, \boldsymbol{\psi}_{l}^{n}\right) \mathbf{v}_{k}\left(\mathbf{u}_{m}\right), \quad k=0, \ldots, K ,
\end{equation}
are the GHs evaluated at the basis obtained via the ``second" DMs embedding $\mathbf{v}_{k}\left(\mathbf{u}_{m}\right)$. This time, the projection coefficients of the function $\boldsymbol{f}\left(\boldsymbol{\psi}_{m}\right)$ to be extended are calculated as $\tilde{a}_{k}=\left\langle\mathbf{v}_{k}, \boldsymbol{f}\left(\boldsymbol{\psi}_{m}\right)\right\rangle$, where $\boldsymbol{\psi}_{m}$ is the DMs embedding of the $m$-th observation. Casting the extension in the matrix form as
\begin{equation}
\mathcal{E}(\hat{\boldsymbol{f}})=\mathbf{K}_{L \times M} \mathbf{V}_{M \times K} \boldsymbol{\Sigma}_{K \times K}^{-1} \mathbf{V}_{K \times M}^{T} \boldsymbol{f}_{M \times 1},
\label{eq:Efcap}
\end{equation}
one obtains a form similar to~(\ref{eq:E2}), where now $\mathbf{V}_{M \times K}$ is the column-wise matrix of the $K$ eigenvectors $\mathbf{v}_{k}$ obtained through the ``second" DMs, and $\boldsymbol{\Sigma}_{K \times K}$ is the diagonal matrix with the corresponding eigenvalues $\sigma_{k}$.

Regarding the numerical solution of the pre-image problem, we are interested in defining the lifting operator for a set of new out-of-sample data points $\boldsymbol{\psi}_{l}^{n} \notin\left\{\boldsymbol{\psi}_{m}=\mathcal{R}\left(\boldsymbol{\omega}_{m}\right) \mid m=1, \ldots, M\right\}$ of the low-dimensional space, compactly written in the matrix form $\boldsymbol{\Psi}^{n} \in \mathbb{R}^{L \times N}$. Thus, substitutions of $\hat{\boldsymbol{f}}$ with $\mathcal{L}$ and of $\boldsymbol{f}$ with $\boldsymbol{\omega}$ (the left and right-hand sides of~(\ref{eq:lifting})) into~(\ref{eq:Efcap}) imply the matrix form
\begin{equation}
\mathcal{L}\left(\boldsymbol{\Psi}^{n}\right)=\mathbf{K}_{L \times M} \mathbf{V}_{M \times K} \boldsymbol{\Sigma}_{K \times K}^{-1} \mathbf{V}_{K \times M}^{T} \boldsymbol{\Omega}_{M \times N},
\label{eq:lifting_2}
\end{equation}
which extends the definition of the lifting operator to reconstruct (pre-image) the $L$ new data points in $\boldsymbol{\Psi}^{n}$ in the high-dimensional space. A schematic representation of the GHs extension is demonstrated in figure~\ref{fig:Diffusion_Maps_meth}.

It is important to note here that the number $K$ of the ``second" DMs eigenvectors needs to be high enough for an accurate reconstruction of the high-dimensional space (see \cite{evangelou2022double} for an indicative algorithm on the selection of $\delta$ threshold). However, the advantage of this method, as evidenced by~(\ref{eq:lifting_2}), is that for the reconstruction of new data points $\boldsymbol{\Psi}^{n}$ only the matrix $\mathbf{K}$ is required. This is because the matrix $\mathbf{V} \Sigma^{-1} \mathbf{V}^{T} \boldsymbol{\Omega}$ is independent of the new data points, and thus it can be pre-computed for accelerating computations (\cite{chiavazzo2014reduced}, \cite{evangelou2022double}).

\subsection{Reconstruction error}
\label{subsec:recon_err}

Once the PDM coordinates have been obtained, we can consider a generic snapshot $\boldsymbol{\omega}_m$ as a new (unseen) data point in the high-dimensional space. By means of the Nyström encoder~(\ref{eq_Nystrom}) and the Geometric Harmonics decoder~(\ref{eq:lifting_2}), we can thus project $\boldsymbol{\omega}_m$ into the low-dimensional manifold, and then lift it back to the high-dimensional space. As a result of the encoding-decoding operators, we obtain the reconstructed vorticity field $\tilde{\boldsymbol{\omega}}_m$, which can be compared with the ``ground truth'' vorticity $\boldsymbol{\omega}_m$. Here, we have chosen to employ the L$_{\infty}$ norm of the difference between $ \boldsymbol{\omega}_m$ and $ \tilde{\boldsymbol{\omega}}_m$ as the comparative metrics, namely $\| \boldsymbol{\omega}_m - \tilde{\boldsymbol{\omega}}_m \|_{\infty}$.

By iteratively applying the procedure described above to all the $M$ collected observations in the high-dimensional space, we can evaluate the reconstruction error distribution over the entire data-set as
\begin{equation}
\label{eq:recon_err}
\varepsilon_m = \dfrac{\| \boldsymbol{\omega}_m - \tilde{\boldsymbol{\omega}}_m \|_{\infty}}{\| \boldsymbol{\omega}_m\|_{\infty}} \times 100, \quad m = 1, \ldots, M ,
\end{equation}
which represents the relative percentage error between the ``ground truth'' solution ($\boldsymbol{\omega}_m$) and the ``reconstruction" value ($\tilde{\boldsymbol{\omega}}_m$). 

The trend of the reconstruction error $\varepsilon_m$ as a function of the Reynolds number $Re$ will be  discussed in \S~\ref{subsec:comparison}, where
the analogous quantity evaluated by means of the Proper Orthogonal Decomposition (POD) will be also reported for comparison. For the sake of completeness, we briefly recall in this Section that the well-known POD technique (\cite{Lumley}) decomposes fluctuations of the stochastic field $\omega(x, y, t)$ with respect to its temporal mean $\overline{\omega}(x,y)$ as
\begin{equation}
\label{eq:POD_def}
\omega^\prime(x, y, t)= \omega(x, y, t)-\overline{\omega}(x)=\sum_{j=1}^{\infty}a^\star_j(t)\varphi^\star_j(x,y),
\end{equation}
in which the POD modes $\varphi^\star_j(x,y)$ are mutually spatially orthogonal. For discrete numerical data, spatio-temporal realizations of the flow field $\omega$ can be organized in the matrix format $\boldsymbol{\Omega} \in \mathbb{R}^{M \times N}$. The discrete POD/PCA modes $\boldsymbol{\varphi}^\star_j$ can then be obtained by employing the method of snapshots by \cite{Sirovich}, namely by solving the eigenvalue problem formulated for the covariance matrix $\mathbf{\Omega}\mathbf{\Omega}^T \in \mathbb{R}^{M \times M}$, reading as
\begin{equation} 
\label{eq:eigprobQQt}
\mathbf{\Omega}\mathbf{\Omega}^T \boldsymbol{\psi}_j^\star=\lambda^\star_j \boldsymbol{\psi}^\star_j, \quad j=1, \ldots, M,
\end{equation}
where $\boldsymbol{\varphi}^\star_j  = \mathbf{\Omega^T} \boldsymbol{\psi}^\star_j/\sqrt{\lambda^\star_j}$ is the $j^{th}$ mode and $\lambda^\star_j$ the corresponding eigenvalue.

\section{Numerical results}
\label{sec:results}

We applied the proposed PDMs algorithm presented in \S~\ref{sec:methods} to the numerical simulation data of the fluidic pinball in the four dynamical regimes outlined in \S~\ref{subsec:simulation}. First, four values of the Reynolds number representative of the different regimes are considered, namely $Re=30$ (\S~\ref{subsec:results-Re_30}), $80$ (\S~\ref{subsec:results-Re_80}), $105$ (\S~\ref{subsec:results-Re_105}) and $130$ (\S~\ref{subsec:results-Re_130}).
%such as to discuss the evolution of the eigenvalues spectrum, the nonlinear embedded manifold, and the flow reconstruction by moving from the vortex shedding symmetric regime to chaotic conditions. 
Later on, the nonlinear PDMs-based reconstruction is compared with a linear POD/PCA-based counterpart for different Reynolds number values (\S~\ref{subsec:comparison}).

For each $Re$, we perform five simulations starting from randomly initialized conditions ((\ref{eq:init}) in \S~\ref{sec:layout}), and we collected 600 observations of the two-dimensional vorticity field from each simulation. Therefore, a data-set of $M=2000$ flow temporal realizations with resolution $\Delta t = 0.5$ is employed to perform the PDMs embedding in each regime.

\subsection{Symmetric periodic regime}
\label{subsec:results-Re_30}

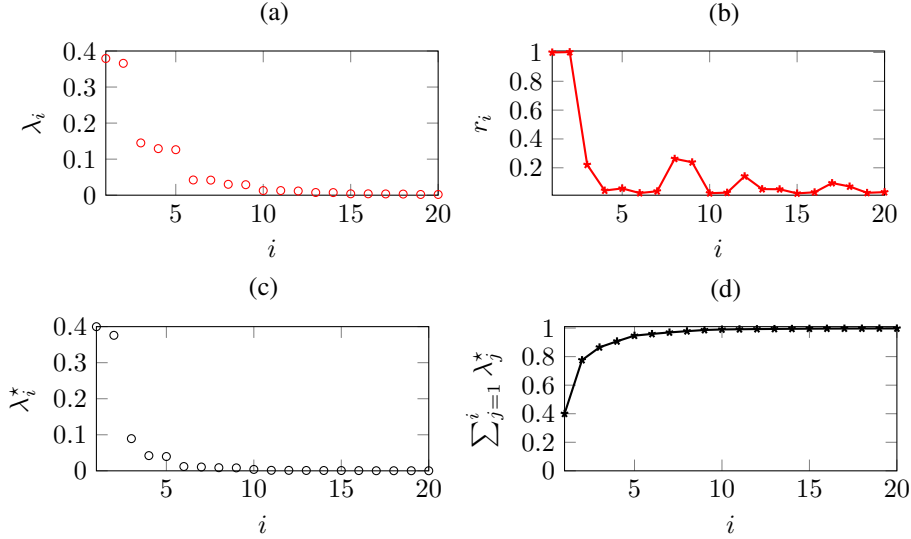
\begin{figure}%[!htb]
	\centering
	\begin{tikzpicture}
	\pgfplotsset{every axis legend/.append style={at={(1.0,1.0)},anchor=north east}}
	\begin{axis}[xlabel=$i$,ylabel=$\lambda_i$,xmin=1,xmax=20,ymin=0,ymax=0.4,width=6.0cm,height=3.5cm]
	\textit{\addplot[only marks,mark=o,mark size=1.5pt,red]
		table [x expr=\thisrowno{0}, y expr=\thisrowno{1}] {DM_eig_and_res_Re_30.dat}; 
		%		\addlegendentry{$Re=30$}
	}
	\end{axis}
	\node[above,yshift=0.0,xshift=0.5cm] at (current bounding box.north) {(a)};
	\end{tikzpicture}
	\begin{tikzpicture}
	\pgfplotsset{every axis legend/.append style={at={(1.0,1.0)},anchor=north east}}
	\begin{axis}[xlabel=$i$,ylabel=$r_i$,xmin=1,xmax=20,ymin=0.01,ymax=1.01,width=6.0cm,height=3.5cm]
\textit{\addplot[thick,black,solid,mark=star,mark size=1.5pt,red]
		table [x expr=\thisrowno{0}, y expr=\thisrowno{2}] {DM_eig_and_res_Re_30.dat}; 
		%		\addlegendentry{$Re=30$}
	}
	\end{axis}
	\node[above,yshift=0.0,xshift=0.5cm] at (current bounding box.north) {(b)};
	\end{tikzpicture}
 	\begin{tikzpicture}
	\pgfplotsset{every axis legend/.append style={at={(1.0,1.0)},anchor=north east}}
	\begin{axis}[xlabel=$i$,ylabel=$\lambda^\star_i$,xmin=1,xmax=20,ymin=0,ymax=0.4,width=6.0cm,height=3.5cm]
	\textit{\addplot[only marks,mark=o,mark size=1.5pt,black]
		table [x expr=\thisrowno{0}, y expr=\thisrowno{1}] {POD_eig_Re_30.dat}; 
		%		\addlegendentry{$Re=30$}
	}
	\end{axis}
	\node[above,yshift=0.0,xshift=0.5cm] at (current bounding box.north) {(c)};
	\end{tikzpicture}
  	\begin{tikzpicture}
	\pgfplotsset{every axis legend/.append style={at={(1.0,1.0)},anchor=north east}}
	\begin{axis}[xlabel=$i$,ylabel=$\sum_{j=1}^{i} \lambda^\star_j$,xmin=1,xmax=20,ymin=0,ymax=1.01,width=6.0cm,height=3.5cm]
 \textit{\addplot[thick,black,solid,mark=star,mark size=1.5pt,black]
		table [x expr=\thisrowno{0}, y expr=\thisrowno{2}] {POD_eig_Re_30.dat}; 
		%		\addlegendentry{$Re=30$}
	}
	\end{axis}
	\node[above,yshift=0.0,xshift=0.5cm] at (current bounding box.north) {(d)};
	\end{tikzpicture}
	\caption{Diffusion maps eigenvalues $\lambda_i$ (a) and corresponding local linear fitting coefficients $r_i$ for the identification of the PDMs (b) as a function of the mode index $i$. In (c)-(d), the corresponding POD/PCA eigenvalues $\lambda^\star_i$ and their cumulative sum are reported, respectively. $Re=30$.
\label{fig:eigenvalues_comparison_Re_30}
	}
\end{figure}

The first twenty DMs are calculated in the symmetric vortex shedding regime for $Re=30$, by selecting a value of the shape parameter equal to $\epsilon = 17.67$, corresponding to $\sim$50\% of the distances being higher than this value. The corresponding eigenvalues are displayed in figure~\ref{fig:eigenvalues_comparison_Re_30}(a), indicating a large spectral gap between the couple $(\lambda_1, \lambda_2)$ and the other eigenvalues. Therefore, the couple of coordinates ($\boldsymbol{\psi_1}$,$\boldsymbol{\psi_2}$) appears to be an adequate choice to parametrize the manifold. This is confirmed by the local linear fitting coefficients of the PDMs distribution $r_i$ reported in figure~\ref{fig:eigenvalues_comparison_Re_30}(b), showing that $r_i \approx 1$ for $i \leq 2$, and then it drops down for $i > 2$. This result indicates that including more than the leading two parsimonious DMs  will not improve the parameterization of the manifold, since their contribution $r_i$ for $i>2$ to the directions along the data set are negligible. Nevertheless, for comparison with previous works of the literature (\cite{Deng_Noack_2020}, \cite{Farzamnik_2023}) and with other cases reported in this section, we select the leading three PDMs coordinates ($\boldsymbol{\psi_1}$,$\boldsymbol{\psi_2}$, $\boldsymbol{\psi_3}$) to represent the low-dimensional subspace embedding the symmetric vortex shedding dynamics in this Reynolds number regime. 

\begin{figure}%[h]
	\centering
	\includegraphics[scale=0.8]{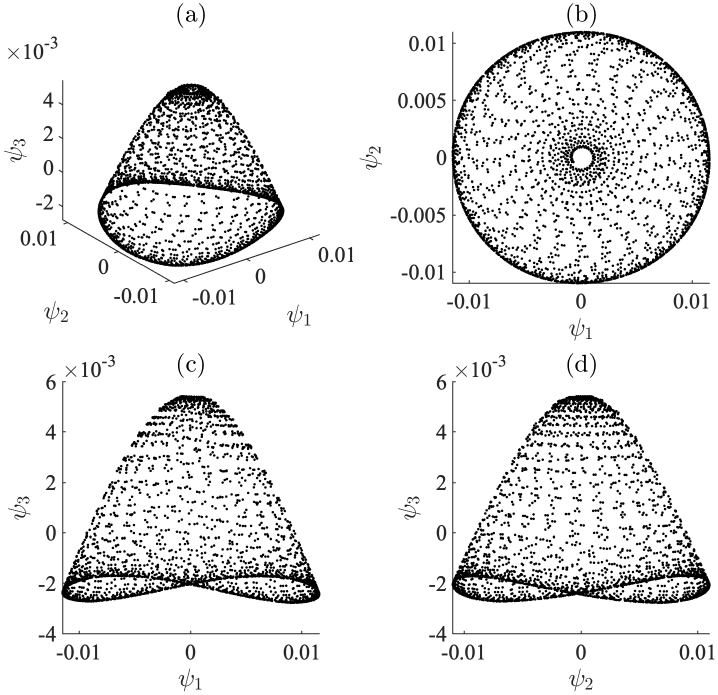}
	\caption{Three-dimensional view (a) and two-dimensional sections ((b)-(c)) of the manifold embedded by the three leading parsimonious diffusion maps (PDMs) coordinates. $Re=30$.
		\label{fig:DM_3D_manifold_Re_30}
	}
\end{figure}

The nonlinear manifold is displayed in figure~\ref{fig:DM_3D_manifold_Re_30} in terms of three-dimensional (a) and two-dimensional views parallel to the planes $\psi_2$-$\psi_1$ (b), $\psi_3$-$\psi_1$ (c) and $\psi_3$-$\psi_2$ (d). The parabolic shape of the manifold first found by \cite{Deng_Noack_2020} is retrieved in figure~\ref{fig:DM_3D_manifold_Re_30}, where both the asymptotic (permanent) limit cycle (external edge of the circle in $\psi_2$-$\psi_1$ plane) and the transient dynamics (inner points of the circle in $\psi_2$-$\psi_1$ plane and points on the paraboloid in $\psi_3$-$\psi_1$ and $\psi_3$-$\psi_2$ planes) can be clearly seen. Moreover, it is interesting to observe that the manifold is perfectly symmetric with respect to the curves $\psi_1 =0$ and $\psi_2 =0$ ((c)-(d), respectively). This is physically related to the symmetry of the base-bleeding jet downstream of the cylinders in this physical regime. As a further representative feature of the underlying symmetric dynamics, note that the manifold in the two planes $\psi_3$-$\psi_1$ and $\psi_3$-$\psi_2$ displays a double-lobe symmetric shape as soon as the circular limit cycle  is reached in the plane $\psi_1$-$\psi_2$ (see also previous figure~\ref{fig:omega_FFT_var_Re}(d) and related discussion in \S~\ref{subsec:simulation}). Therefore, we can conclude that the two leading PDMs coordinates, namely $\boldsymbol{\psi_1}$ and $ \boldsymbol{\psi_2}$, are physically related to the periodic von Karman street of vortices establishing in the long-time limit downstream of the cylinders. On the other hand, the third coordinate $\boldsymbol{\psi_3}$ is associated to a shift-mode characteristic of the transient dynamics from the onset of vortex shedding to the periodic von Karman wake, analogously to what was found by \cite{Farzamnik_2023}. In this physical regime, the transient dynamics has a minor importance ($r_3$ small with respect to $r_1$ and $r_2$, see again figure~\ref{fig:eigenvalues_comparison_Re_30}(b)), the flow field quick approaching the symmetric limit cycle within the $\psi_1$-$\psi_2$ plane.

\begin{figure}%[h]
	\centering
	\includegraphics[scale=0.8]{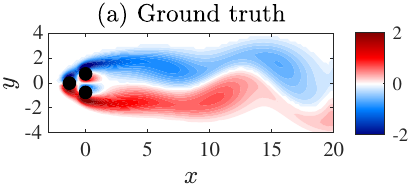}
	\includegraphics[scale=0.8]{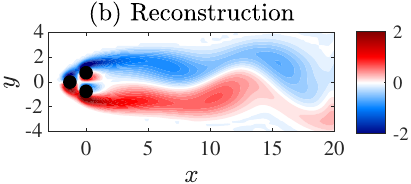}
	\includegraphics[scale=0.8]{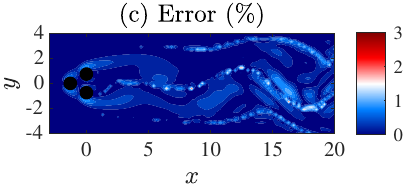}
	\caption{\label{fig:omega_contour_recon_Re_30} Snapshot of the vorticity field $\omega(x,y)$ from the high-dimensional simulation (ground truth solution, (a)) compared with the low-dimensional reconstruction by means of the three leading PDMs coordinates (b). The relative percentage spread is also reported in (c). $Re=30$.}
\end{figure}

The three leading parsimonious diffusion maps  coordinates are employed to reconstruct the vorticity field at each time instant, following the procedure previously described in \S~\ref{subsec:recon_err}. The comparison between the snapshot ground truth solution $\omega_m(x,y)$ and the corresponding reconstruction $\tilde{\omega}_m(x,y)$ for $m = 1000$ is reported in figure~\ref{fig:omega_contour_recon_Re_30}(a)-(b). The difference between the two instantaneous vorticity fields (normalized with respect to the ground truth solution) is shown in figure~\ref{fig:omega_contour_recon_Re_30}(c), while a more comprehensive representation of the DMs-based reconstruction performance is given in Table~\ref{tab:Re_30_recon_error_DM_GH}, which reports the mean value and the percentiles (5$^{th}$, 10$^{th}$, 50$^{th}$, 90$^{th}$ and 95$^{th}$) of the relative percentage error $\varepsilon_m$ (see~(\ref{eq:recon_err}) in \S~\ref{subsec:recon_err}) evaluated over the entire data-set. In particular, it can be seen that both the mean and the 95$^{th}$ percentile of the error distribution are below 3$\%$.

\begin{table}
\begin{center}
%	\vspace{0.2cm}
	%	\begin{tabular}{lcccccc}
	%		\hline
	%		\textbf{Percentile} & \textbf{RMSE} & \textbf{L1} & \textbf{L2} & \textbf{L$\infty$} \\
	%		\hline
	%5  & 1.57 & 1.32 & 1.57 & 1.74 \\
	%10 & 1.63 & 1.40 & 1.63 & 1.80 \\
	%50 & 1.93 & 1.99 & 1.93 & 2.14 \\
	%90 & 2.71 & 2.94 & 2.71 & 2.64 \\
	%95 & 2.92 & 3.16 & 2.92 & 2.87 \\
	%		\hline
	%		\textbf{Mean} & \textbf{RMSE} & \textbf{L1} & \textbf{L2} & \textbf{L$\infty$}\\
	%		\hline
	%        -- & 2.05 & 2.05 & 2.05 & 2.19 \\
	%	\end{tabular}
	\renewcommand{\arraystretch}{1.5} % Adjust the value as needed
	\begin{tabular}{lcccccc}
	%	\hline
		\textbf{Percentile} & 	5  & 	10 & 	50 & 	90 & 	95 &\\
		\hline
		& 1.74 \%& 1.80 \%& 2.14 \%& 2.64 \%& 2.87\% \\
		\hline
		\textbf{Mean value} & 2.19 \%\\
		%\hline
	\end{tabular}
\end{center}
		\caption{Distribution of the relative percentage error $\varepsilon_m$ between the ground truth solution and the reconstruction with three leading PDMs coordinates over the entire data-set. Reynolds number $Re=30$.}
\label{tab:Re_30_recon_error_DM_GH}
\end{table}

\subsection{Asymmetric periodic regime}
\label{subsec:results-Re_80}

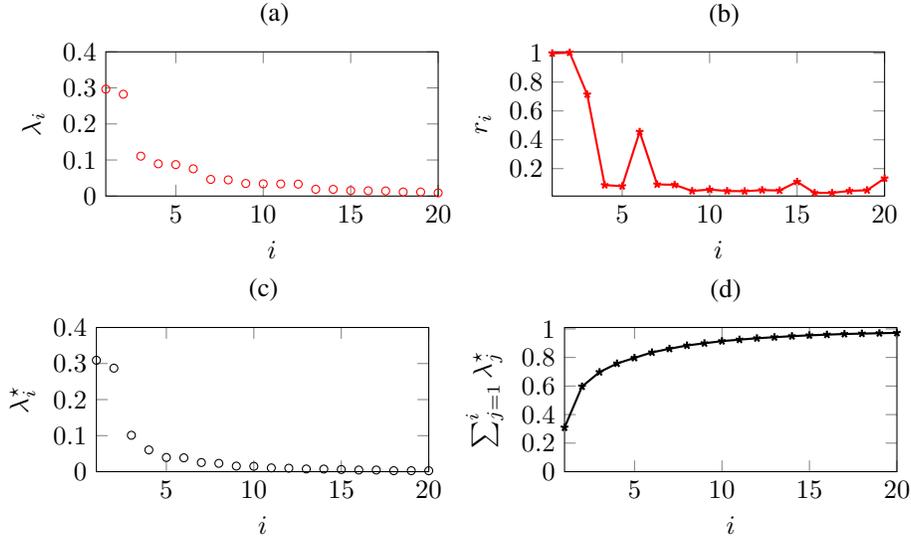
\begin{figure}%[!htb]
	\centering
	\begin{tikzpicture}
	\pgfplotsset{every axis legend/.append style={at={(1.0,1.0)},anchor=north east}}
	\begin{axis}[xlabel=$i$,ylabel=$\lambda_i$,xmin=1,xmax=20,ymin=0,ymax=0.4,width=6.0cm,height=3.5cm]
	\textit{\addplot[only marks,mark=o,mark size=1.5pt,red]
		table [x expr=\thisrowno{0}, y expr=\thisrowno{1}] {DM_eig_and_res_Re_80.dat}; 
		%		\addlegendentry{$Re=30$}
	}
	\end{axis}
	\node[above,yshift=0.0,xshift=0.5cm] at (current bounding box.north) {(a)};
	\end{tikzpicture}
	\begin{tikzpicture}
	\pgfplotsset{every axis legend/.append style={at={(1.0,1.0)},anchor=north east}}
	\begin{axis}[xlabel=$i$,ylabel=$r_i$,xmin=1,xmax=20,ymin=0.01,ymax=1.01,width=6.0cm,height=3.5cm]
\textit{\addplot[thick,black,solid,mark=star,mark size=1.5pt,red]
		table [x expr=\thisrowno{0}, y expr=\thisrowno{2}] {DM_eig_and_res_Re_80.dat}; 
		%		\addlegendentry{$Re=30$}
	}
	\end{axis}
	\node[above,yshift=0.0,xshift=0.5cm] at (current bounding box.north) {(b)};
	\end{tikzpicture}
 	\begin{tikzpicture}
	\pgfplotsset{every axis legend/.append style={at={(1.0,1.0)},anchor=north east}}
	\begin{axis}[xlabel=$i$,ylabel=$\lambda^\star_i$,xmin=1,xmax=20,ymin=0,ymax=0.4,width=6.0cm,height=3.5cm]
	\textit{\addplot[only marks,mark=o,mark size=1.5pt,black]
		table [x expr=\thisrowno{0}, y expr=\thisrowno{1}] {POD_eig_Re_80.dat}; 
		%		\addlegendentry{$Re=30$}
	}
	\end{axis}
	\node[above,yshift=0.0,xshift=0.5cm] at (current bounding box.north) {(c)};
	\end{tikzpicture}
  	\begin{tikzpicture}
	\pgfplotsset{every axis legend/.append style={at={(1.0,1.0)},anchor=north east}}
	\begin{axis}[xlabel=$i$,ylabel=$\sum_{j=1}^{i} \lambda^\star_j$,xmin=1,xmax=20,ymin=0,ymax=1.01,width=6.0cm,height=3.5cm]
 \textit{\addplot[thick,black,solid,mark=star,mark size=1.5pt,black]
		table [x expr=\thisrowno{0}, y expr=\thisrowno{2}] {POD_eig_Re_80.dat}; 
		%		\addlegendentry{$Re=30$}
	}
	\end{axis}
	\node[above,yshift=0.0,xshift=0.5cm] at (current bounding box.north) {(d)};
	\end{tikzpicture}
	\caption{Diffusion maps eigenvalues $\lambda_i$ (a) and corresponding local linear fitting coefficients $r_i$ for the identification of the PDMs (b) as a function of the mode index $i$. In (c)-(d), the corresponding POD/PCA eigenvalues $\lambda^\star_i$ and their cumulative sum are reported, respectively. $Re=80$.
\label{fig:eigenvalues_comparison_Re_80}
	}
\end{figure}

The first twenty DMs are calculated in the asymmetric vortex shedding regime for $Re=80$, by selecting a value of the shape parameter equal to $\epsilon = 23.98$. The corresponding eigenvalues are displayed in figure~\ref{fig:eigenvalues_comparison_Re_80}(a), indicating again a large spectral gap between the couple $(\lambda_1, \lambda_2)$ and the other eigenvalues. On the other hand, if one looks at the local linear fitting coefficients distribution $r_i$ of the PDMs, reported in figure~\ref{fig:eigenvalues_comparison_Re_80}(b), it can be seen that the value for $i=3$ ($r_3 \approx 0.8$) is comparable with $r_1 \approx r_2 \approx 1$. Therefore, the parsimonious diffusion maps algorithm detects in this regime a manifold with three latent dimensions, which are represented by the triple of parsimonious diffusion maps coordinates ($\boldsymbol{\psi_1}$,$\boldsymbol{\psi_2}$,$\boldsymbol{\psi_3}$).

\begin{figure}%[h]
	\centering
	\includegraphics[scale=0.8]{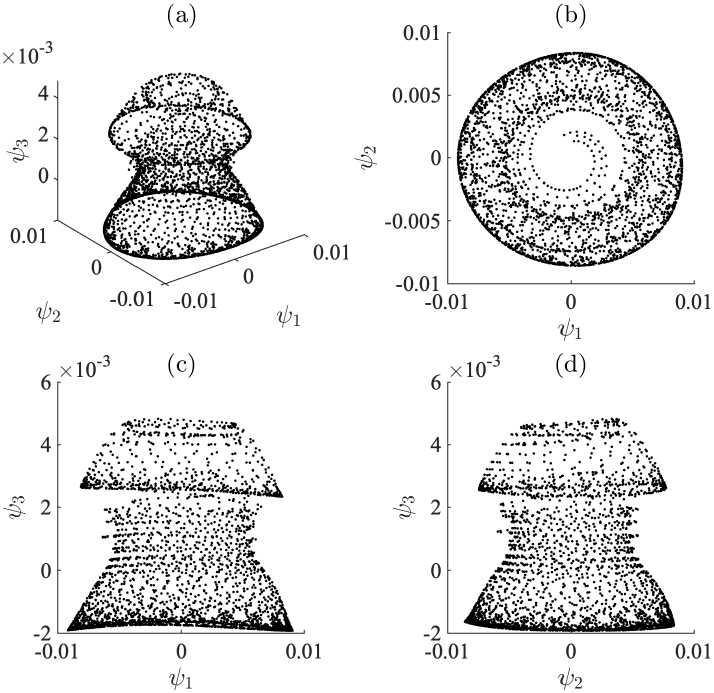}
	\caption{Three-dimensional view (a) and two-dimensional sections ((b)-(c)) of the manifold embedded by the three leading parsimonious diffusion maps coordinates. $Re=80$.
		\label{fig:DM_3D_manifold_Re_80}
	}
\end{figure}

The nonlinear manifold is displayed in  figure~\ref{fig:DM_3D_manifold_Re_80} in terms of three-dimensional (a) and two-dimensional views parallel to the planes $\psi_2$-$\psi_1$ (b), $\psi_3$-$\psi_1$ (c) and $\psi_3$-$\psi_2$ (d). To give a physical interpretation of the manifold shape, it is useful to recall that, in this Reynolds number regime, the system has undergone a pitchfork bifurcation (see discussion in the introduction and \cite{Deng_Noack_2020} for further details). As a result, two additional unstable steady solutions occur, breaking the reflection symmetry of the flow configuration represented by the symmetric solution already existing from the previous regime. The asymmetry of the new solutions reflects into the mean field, which displays an upward or downward-deflected base-bleeding jet right downstream of the cylinders array. As reported in \cite{Deng_Noack_2020}, the statistically symmetric limit cycle, which is associated with the statistically symmetric vortex shedding, becomes unstable with respect to two mirror-conjugated statistically asymmetric limit cycles, associated with statistically asymmetric von Karman street of vortices. The existence of these multiple solutions, namely one unstable symmetric and two mirror-conjugated asymmetric limit cycles, can be retrieved in the ``mushroom" shape of the manifold reported in figure~\ref{fig:DM_3D_manifold_Re_80}(a). More in detail, the two asymmetric limit cycles associated with the upward and downward-deflected shedding of vortices are represented by two superposed circular edges of points in the $\psi_2$-$\psi_1$ plane (figure~\ref{fig:DM_3D_manifold_Re_80}(b)). On the other hand, the clouds of points in the planes $\psi_3$-$\psi_1$ (c) and $\psi_3$-$\psi_2$ (d) are again related to the transient dynamics towards the limit cycles. In this respect, it is interesting to observe that the manifold embeds the transient exploration of the two stable asymmetric solutions as well as of the unstable symmetric flow pattern, which finally converges towards one of the two co-existing attractors in the state space (i.e. the circular edges of the manifold for $\psi_3 > 0$ and $\psi_3 < 0$). Note that analogous considerations were reported by \cite{Deng_Noack_2021} when discussing the trajectories originating from different (symmetric and asymmetric) initial conditions in the lift-drag plane for $Re=80$ (see in particular figure~11 by \cite{Deng_Noack_2021}).

\begin{figure}%[h]
	\centering
	\includegraphics[scale=0.8]{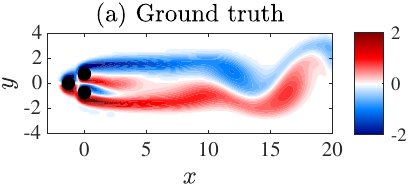}
	\includegraphics[scale=0.8]{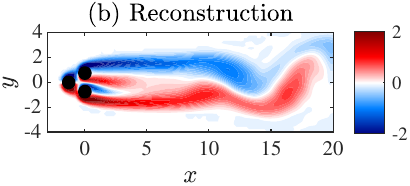}
	\includegraphics[scale=0.8]{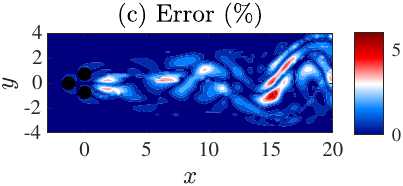}
	\caption{\label{fig:omega_contour_recon_Re_80} Snapshot of the vorticity field $\omega(x,y)$ from the high-dimensional simulation (ground truth solution, (a)) compared with the low-dimensional reconstruction by means of the three leading PDMs coordinates (b). The relative percentage error is also reported in (c). $Re=80$.}
\end{figure}

\begin{table}
	\begin{center}
		%	\vspace{0.2cm}
		%	\begin{tabular}{lcccccc}
		%		\hline
		%		\textbf{Percentile} & \textbf{RMSE} & \textbf{L1} & \textbf{L2} & \textbf{L$\infty$} \\
		%		\hline
		%5  & 1.57 & 1.32 & 1.57 & 1.74 \\
		%10 & 1.63 & 1.40 & 1.63 & 1.80 \\
		%50 & 1.93 & 1.99 & 1.93 & 2.14 \\
		%90 & 2.71 & 2.94 & 2.71 & 2.64 \\
		%95 & 2.92 & 3.16 & 2.92 & 2.87 \\
		%		\hline
		%		\textbf{Mean} & \textbf{RMSE} & \textbf{L1} & \textbf{L2} & \textbf{L$\infty$}\\
		%		\hline
		%        -- & 2.05 & 2.05 & 2.05 & 2.19 \\
		%	\end{tabular}
		\renewcommand{\arraystretch}{1.5} % Adjust the value as needed
		\begin{tabular}{lcccccc}
			\hline
			\textbf{Percentile} & 	5  & 	10 & 	50 & 	90 & 	95 &\\
			\hline
			& 2.27 \%& 2.63 \%& 5.13 \%& 10.77 \%& 15.84\% \\
			\hline
			\textbf{Mean value} & 6.33 \%\\
			\hline
		\end{tabular}
	\end{center}
	\caption{Distribution of the relative percentage error $\varepsilon_m$ between the ground truth solution and the reconstruction with three leading parsimonious diffusion maps (PDMs) coordinates over the entire data-set. Reynolds number $Re=80$.}
	\label{tab:Re_80_recon_error_DM_GH}
\end{table}

The three leading PDMs coordinates are employed to reconstruct the vorticity field at each time instant, following the procedure previously described in \S~\ref{subsec:recon_err}. The comparison between the snapshot ground truth solution $\omega_m(x,y)$ and the corresponding reconstruction $\tilde{\omega}_m(x,y)$ for $m = 1000$ is reported in figure~\ref{fig:omega_contour_recon_Re_80}(a)-(b). The difference between the two instantaneous vorticity fields (normalized with respect to the ground truth solution) is shown in figure~\ref{fig:omega_contour_recon_Re_80}(c), while a more comprehensive representation of the PDMs-based reconstruction performance is given in Table~\ref{tab:Re_80_recon_error_DM_GH}, which reports the mean value and the percentiles (5$^{th}$, 10$^{th}$, 50$^{th}$, 90$^{th}$ and 95$^{th}$) of the relative percentage error $\varepsilon_m$ (see~(\ref{eq:recon_err}) in \S~\ref{subsec:recon_err}) evaluated over the entire data-set. In particular, it can be seen that the 95$^{th}$ percentile of the error distribution is approximately 15\%, while the mean value is below $7 \%$.

\subsection{Asymmetric quasi-periodic regime}
\label{subsec:results-Re_105}

\begin{figure}%[!htb]
	\centering
	\begin{tikzpicture}
	\pgfplotsset{every axis legend/.append style={at={(1.0,1.0)},anchor=north east}}
	\begin{axis}[xlabel=$i$,ylabel=$\lambda_i$,xmin=1,xmax=20,ymin=0,ymax=0.4,width=6.0cm,height=3.5cm]
	\textit{\addplot[only marks,mark=o,mark size=1.5pt,red]
		table [x expr=\thisrowno{0}, y expr=\thisrowno{1}] {DM_eig_and_res_Re_105.dat}; 
		%		\addlegendentry{$Re=30$}
	}
	\end{axis}
	\node[above,yshift=0.0,xshift=0.5cm] at (current bounding box.north) {(a)};
	\end{tikzpicture}
	\begin{tikzpicture}
	\pgfplotsset{every axis legend/.append style={at={(1.0,1.0)},anchor=north east}}
	\begin{axis}[xlabel=$i$,ylabel=$r_i$,xmin=1,xmax=20,ymin=0.01,ymax=1.01,width=6.0cm,height=3.5cm]
\textit{\addplot[thick,black,solid,mark=star,mark size=1.5pt,red]
		table [x expr=\thisrowno{0}, y expr=\thisrowno{2}] {DM_eig_and_res_Re_105.dat}; 
		%		\addlegendentry{$Re=30$}
	}
	\end{axis}
	\node[above,yshift=0.0,xshift=0.5cm] at (current bounding box.north) {(b)};
	\end{tikzpicture}
 	\begin{tikzpicture}
	\pgfplotsset{every axis legend/.append style={at={(1.0,1.0)},anchor=north east}}
	\begin{axis}[xlabel=$i$,ylabel=$\lambda^\star_i$,xmin=1,xmax=20,ymin=0,ymax=0.4,width=6.0cm,height=3.5cm]
	\textit{\addplot[only marks,mark=o,mark size=1.5pt,black]
		table [x expr=\thisrowno{0}, y expr=\thisrowno{1}] {POD_eig_Re_105.dat}; 
		%		\addlegendentry{$Re=30$}
	}
	\end{axis}
	\node[above,yshift=0.0,xshift=0.5cm] at (current bounding box.north) {(c)};
	\end{tikzpicture}
  	\begin{tikzpicture}
	\pgfplotsset{every axis legend/.append style={at={(1.0,1.0)},anchor=north east}}
	\begin{axis}[xlabel=$i$,ylabel=$\sum_{j=1}^{i} \lambda^\star_j$,xmin=1,xmax=20,ymin=0,ymax=1.01,width=6.0cm,height=3.5cm]
 \textit{\addplot[thick,black,solid,mark=star,mark size=1.5pt,black]
		table [x expr=\thisrowno{0}, y expr=\thisrowno{2}] {POD_eig_Re_105.dat}; 
		%		\addlegendentry{$Re=30$}
	}
	\end{axis}
	\node[above,yshift=0.0,xshift=0.5cm] at (current bounding box.north) {(d)};
	\end{tikzpicture}
	\caption{Diffusion Maps eigenvalues $\lambda_i$ (a) and corresponding local linear fitting coefficients $r_i$ for the identification of the PDMs (b) as a function of the mode index $i$. In (c)-(d), the corresponding POD eigenvalues $\lambda^\star_i$ and their cumulative sum are reported, respectively. $Re=105$.
\label{fig:eigenvalues_comparison_Re_105}
	}
\end{figure}
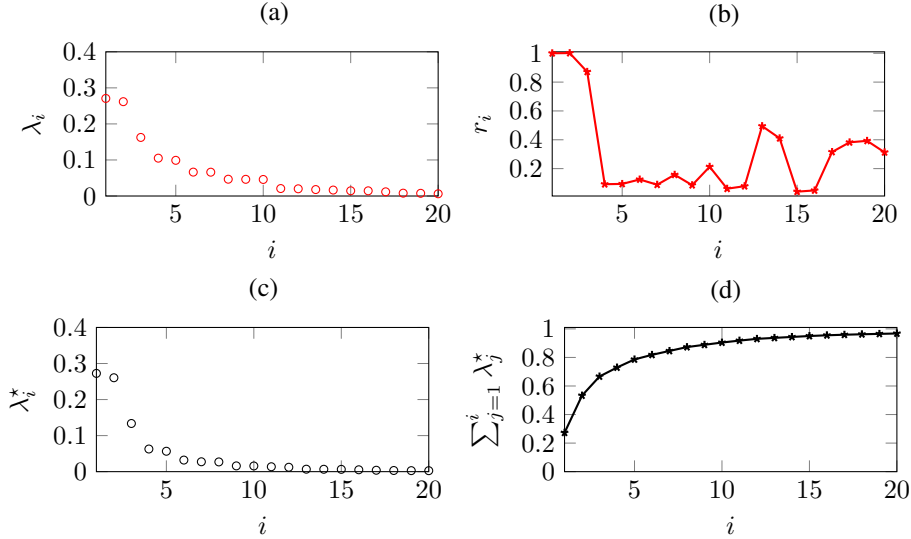

The first twenty DMs are calculated in the quasi-periodic asymmetric vortex shedding regime for $Re=105$, by selecting a value of the shape parameter equal to for DP: $\epsilon = 31.74$. The corresponding eigenvalues are displayed in figure~\ref{fig:eigenvalues_comparison_Re_105}(a), while the local linear fitting coefficients distribution $r_i$ for the identification of the PDMs, is reported in figure~\ref{fig:eigenvalues_comparison_Re_105}(b). By looking at the latter panel, it can be clearly appreciated that the important latent dimensions where the flow dynamics evolves are again three: $r_i \approx 1$ for $i \leq 3$, and then it drops down for $i > 3$. Therefore, the triple of coordinates ($\boldsymbol{\psi_1}$,$\boldsymbol{\psi_2}$,$\boldsymbol{\psi_3}$) appears again to be an adequate choice to parametrize the manifold.

\begin{figure}%[h]
	\centering
	\includegraphics[scale=0.8]{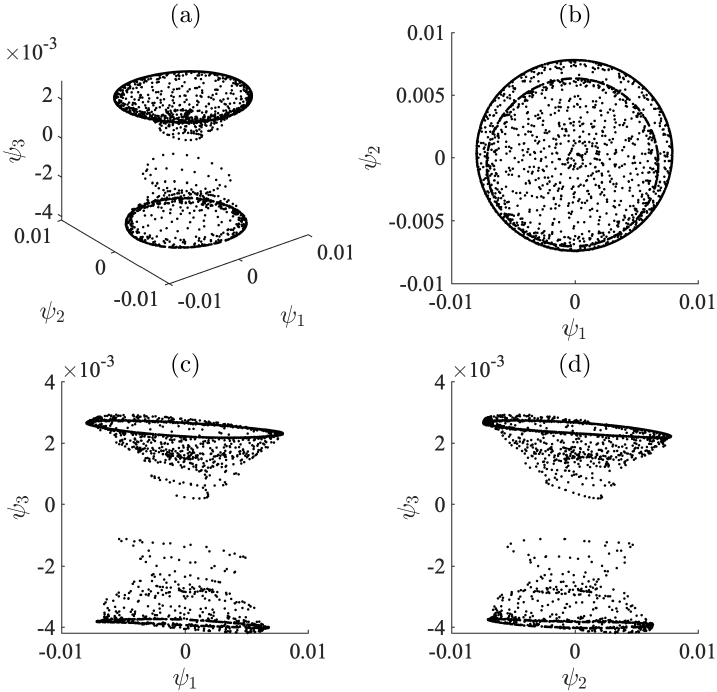}
	\caption{Three-dimensional view (a) and two-dimensional sections ((b)-(c)) of the manifold embedded by the three leading parsimonious diffusion maps coordinates. $Re=105$.
		\label{fig:DM_3D_manifold_Re_105}
	}
\end{figure}

The nonlinear manifold is displayed in  figure~\ref{fig:DM_3D_manifold_Re_105} in terms of three-dimensional (a) and two-dimensional views parallel to the planes $\psi_2$-$\psi_1$ (b), $\psi_3$-$\psi_1$ (c) and $\psi_3$-$\psi_2$ (d). To give a physical interpretation of the manifold shape, it is again useful to recall that two asymmetric limit cycles co-exist as state-space attractors of the flow dynamics also in this regime. However, as pointed out by \cite{Deng_Noack_2021, Deng_Noack_2022}, the flow does not exhibit anymore symmetric oscillations. As a matter of fact, the transient dynamics from the symmetric (unstable) steady solution first reaches one of the two asymmetric (stable) steady configurations, and then evolves towards the corresponding attracting limit cycle (see figure~16 of \cite{Deng_Noack_2021} and figure~4 of \cite{Deng_Noack_2022}). The last considerations help us in understanding the ``hourglass" shape of the manifold obtained at $Re=105$, especially when comparing it to the $Re=80$ case (see previous figure~\ref{fig:DM_3D_manifold_Re_80} and related discussion in \S~\ref{subsec:results-Re_80}). In particular, the manifold obtained for $Re=105$ appears to be separated into two clusters of points (figure~\ref{fig:DM_3D_manifold_Re_105}(a)), thus loosing the continuous ``mushroom" shape found for $Re=80$. By looking at the $\psi_2$-$\psi_1$ plane distribution (b), it can be clearly appreciated that these clusters are physically related to the two co-existing attractors and the corresponding transient dynamical regimes, namely the two asymmetric (for $\psi_3 > 0$ and $\psi_3 < 0$) von Karman street of vortices. The net separation of the two clusters in the planes $\psi_3$-$\psi_1$ (c) and $\psi_3$-$\psi_2$ (d) denotes the absence of symmetric (transient or post-transient) oscillations in the current flow regime.

\begin{figure}%[h]
	\centering
	\includegraphics[scale=0.8]{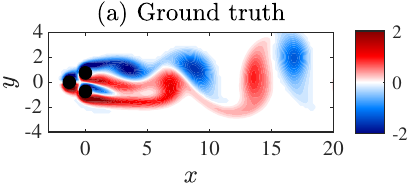}
	\includegraphics[scale=0.8]{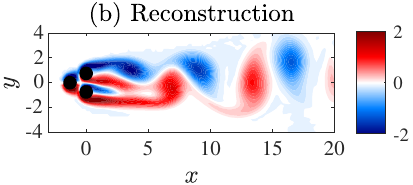}
	\includegraphics[scale=0.8]{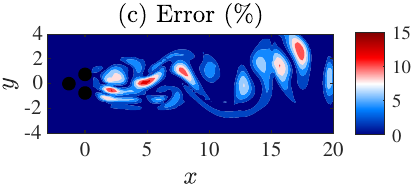}
	\caption{\label{fig:omega_contour_recon_Re_105} Snapshot of the vorticity field $\omega(x,y)$ from the high-dimensional simulation (ground truth solution, (a)) compared with the low-dimensional reconstruction by means of three leading PDMs coordinates (b). The relative percentage error is also reported in (c). $Re=105$.}
\end{figure}

\begin{table}
	\begin{center}
		%	\vspace{0.2cm}
		%	\begin{tabular}{lcccccc}
		%		\hline
		%		\textbf{Percentile} & \textbf{RMSE} & \textbf{L1} & \textbf{L2} & \textbf{L$\infty$} \\
		%		\hline
		%5  & 1.57 & 1.32 & 1.57 & 1.74 \\
		%10 & 1.63 & 1.40 & 1.63 & 1.80 \\
		%50 & 1.93 & 1.99 & 1.93 & 2.14 \\
		%90 & 2.71 & 2.94 & 2.71 & 2.64 \\
		%95 & 2.92 & 3.16 & 2.92 & 2.87 \\
		%		\hline
		%		\textbf{Mean} & \textbf{RMSE} & \textbf{L1} & \textbf{L2} & \textbf{L$\infty$}\\
		%		\hline
		%        -- & 2.05 & 2.05 & 2.05 & 2.19 \\
		%	\end{tabular}
		\renewcommand{\arraystretch}{1.5} % Adjust the value as needed
	\begin{tabular}{lcccccc}
	\hline
	\textbf{Percentile} & 	5  & 	10 & 	50 & 	90 & 	95 &\\
	\hline
	& 3.10 \%& 3.27 \%& 4.52 \%& 21.83 \%& 24.71 \% \\
	\hline
	\textbf{Mean value} & 8.91 \%\\
	\hline
\end{tabular}
	\end{center}
	\caption{Distribution of the relative percentage error $\varepsilon_m$ between the ground truth solution and the reconstruction with three leading parsimonious diffusion maps coordinates over the entire data-set. Reynolds number $Re=105$.}
	\label{tab:Re_105_recon_error_DM_GH}
\end{table}

The three leading parsimonious diffusion maps coordinates are employed to reconstruct the vorticity field at each time instant, following the procedure previously described in \S~\ref{subsec:recon_err}. The comparison between the snapshot ground truth solution $\omega_m(x,y)$ and the corresponding reconstruction $\tilde{\omega}_m(x,y)$ for $m = 1000$ is reported in figure~\ref{fig:omega_contour_recon_Re_105}(a)-(b). The difference between the two instantaneous vorticity fields (normalized with respect to the ground truth solution) is shown in figure~\ref{fig:omega_contour_recon_Re_105}(c), while a more comprehensive representation of the PDMs-based reconstruction performance is given in Table~\ref{tab:Re_105_recon_error_DM_GH}, which reports the mean value and the percentiles (5$^{th}$, 10$^{th}$, 50$^{th}$, 90$^{th}$ and 95$^{th}$) of the relative percentage error $\varepsilon_m$ (see~(\ref{eq:recon_err}) in \S~\ref{subsec:recon_err}) evaluated over the entire data-set. In particular, it can be seen that the 95$^{th}$ percentile of the error distribution is approximately 24\%, while the mean value is below $9 \%$.

\subsection{Chaotic regime}
\label{subsec:results-Re_130}

\begin{figure}%[!htb]
	\centering
	\begin{tikzpicture}
	\pgfplotsset{every axis legend/.append style={at={(1.0,1.0)},anchor=north east}}
	\begin{axis}[xlabel=$i$,ylabel=$\lambda_i$,xmin=1,xmax=20,ymin=0,ymax=0.4,width=6.0cm,height=3.5cm]
	\textit{\addplot[only marks,mark=o,mark size=1.5pt,red]
		table [x expr=\thisrowno{0}, y expr=\thisrowno{1}] {DM_eig_and_res_Re_130.dat}; 
		%		\addlegendentry{$Re=30$}
	}
	\end{axis}
	\node[above,yshift=0.0,xshift=0.5cm] at (current bounding box.north) {(a)};
	\end{tikzpicture}
	\begin{tikzpicture}
	\pgfplotsset{every axis legend/.append style={at={(1.0,1.0)},anchor=north east}}
	\begin{axis}[xlabel=$i$,ylabel=$r_i$,xmin=1,xmax=20,ymin=0.01,ymax=1.01,width=6.0cm,height=3.5cm]
\textit{\addplot[thick,black,solid,mark=star,mark size=1.5pt,red]
		table [x expr=\thisrowno{0}, y expr=\thisrowno{2}] {DM_eig_and_res_Re_130.dat}; 
		%		\addlegendentry{$Re=30$}
	}
	\end{axis}
	\node[above,yshift=0.0,xshift=0.5cm] at (current bounding box.north) {(b)};
	\end{tikzpicture}
 	\begin{tikzpicture}
	\pgfplotsset{every axis legend/.append style={at={(1.0,1.0)},anchor=north east}}
	\begin{axis}[xlabel=$i$,ylabel=$\lambda^\star_i$,xmin=1,xmax=20,ymin=0,ymax=0.4,width=6.0cm,height=3.5cm]
	\textit{\addplot[only marks,mark=o,mark size=1.5pt,black]
		table [x expr=\thisrowno{0}, y expr=\thisrowno{1}] {POD_eig_Re_130.dat}; 
		%		\addlegendentry{$Re=30$}
	}
	\end{axis}
	\node[above,yshift=0.0,xshift=0.5cm] at (current bounding box.north) {(c)};
	\end{tikzpicture}
  	\begin{tikzpicture}
	\pgfplotsset{every axis legend/.append style={at={(1.0,1.0)},anchor=north east}}
	\begin{axis}[xlabel=$i$,ylabel=$\sum_{j=1}^{i} \lambda^\star_j$,xmin=1,xmax=20,ymin=0,ymax=1.01,width=6.0cm,height=3.5cm]
 \textit{\addplot[thick,black,solid,mark=star,mark size=1.5pt,black]
		table [x expr=\thisrowno{0}, y expr=\thisrowno{2}] {POD_eig_Re_130.dat}; 
		%		\addlegendentry{$Re=30$}
	}
	\end{axis}
	\node[above,yshift=0.0,xshift=0.5cm] at (current bounding box.north) {(d)};
	\end{tikzpicture}
	\caption{Diffusion Maps eigenvalues $\lambda_i$ (a) and corresponding local linear fitting coefficients $r_i$ for the identification of the PDMs (b) as a function of the mode index $i$. In (c)-(d), the corresponding POD eigenvalues $\lambda^\star_i$ and their cumulative sum are reported, respectively. $Re=130$.
\label{fig:eigenvalues_comparison_Re_130}
	}
\end{figure}
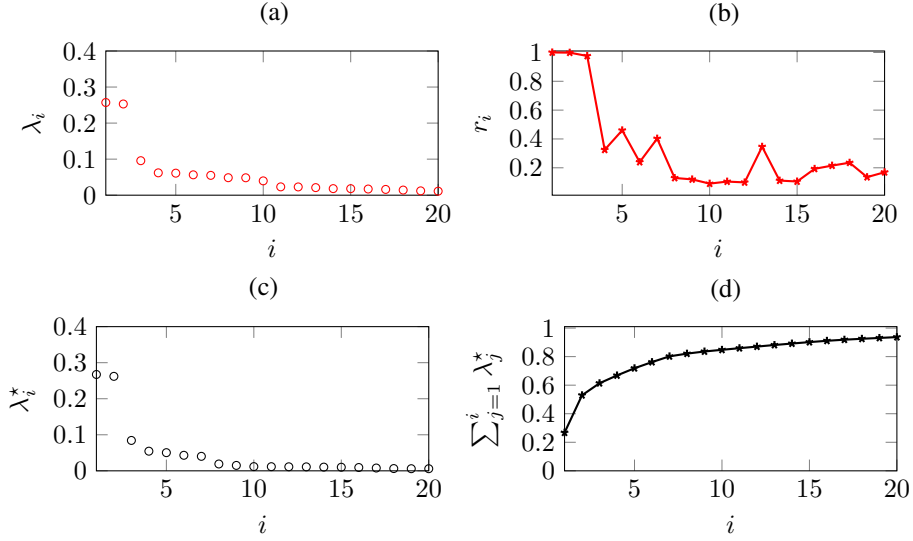

The first twenty DMs are calculated in the chaotic (statistically) symmetric regime for $Re=130$, by selecting a value of the shape parameter equal to  $\epsilon = 33.97$. The corresponding eigenvalues are displayed in figure~\ref{fig:eigenvalues_comparison_Re_130}(a), while the local linear fitting coefficients distribution $r_i$ used for the identification of the PDMs is reported in figure~\ref{fig:eigenvalues_comparison_Re_130}(b). By looking at the latter panel, it can be clearly appreciated that the important latent dimensions where the flow dynamics evolves are three also in this regime: $r_i \approx 1$ for $i \leq 3$, and then it drops down for $i > 3$. Therefore, the triple of coordinates ($\boldsymbol{\psi_1}$, $\boldsymbol{\psi_2}$, $\boldsymbol{\psi_3}$)  appears again to be an adequate choice to parametrize the manifold.

\begin{figure}%[h]
	\centering
	\includegraphics[scale=0.8]{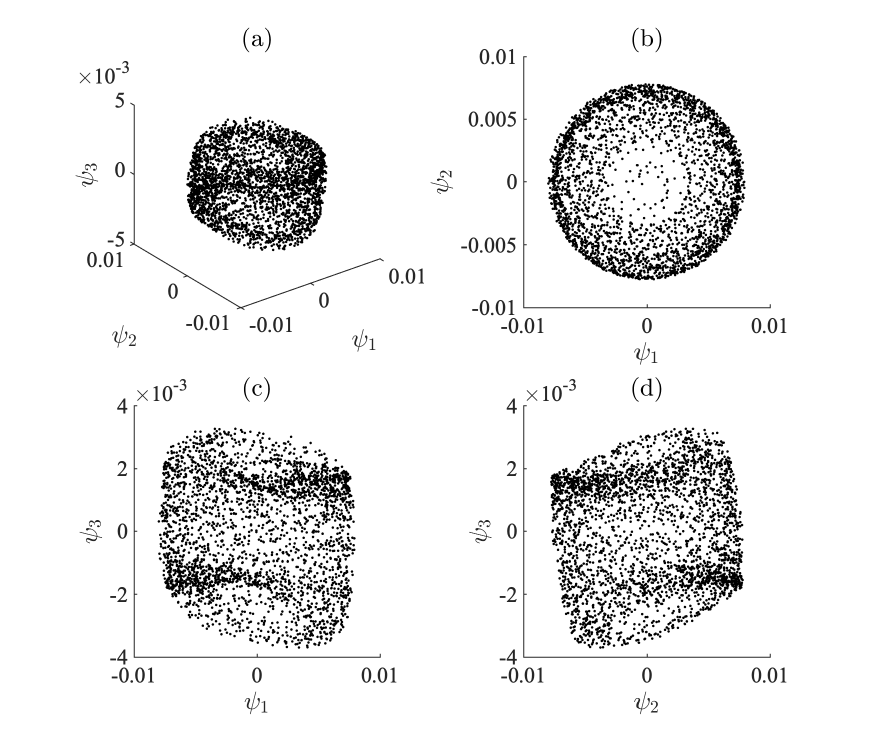}
	\caption{Three-dimensional view (a) and two-dimensional sections ((b)-(c)) of the manifold embedded by the three leading Diffusion Maps coordinates. $Re=130$.
		\label{fig:DM_3D_manifold_Re_130}
	}
\end{figure}

The nonlinear manifold is displayed in  figure~\ref{fig:DM_3D_manifold_Re_130} in terms of three-dimensional (a) and two-dimensional views parallel to the planes $\psi_2$-$\psi_1$ (b), $\psi_3$-$\psi_1$ (c) and $\psi_3$-$\psi_2$ (d). Although the periodic von Karman street of vortices can still be recognized as the dense circular edge of points in the $\psi_2$-$\psi_1$ view of the manifold (b), all panels of figure~\ref{fig:DM_3D_manifold_Re_130} reveal the (statistically symmetric) chaotic nature of the pinball flow in this regime. Differently from the cases analyzed so far, the attractor is a fully three-dimensional object, thus filling the whole state-space.

\begin{figure}%[h]
	\centering
	\includegraphics[scale=0.8]{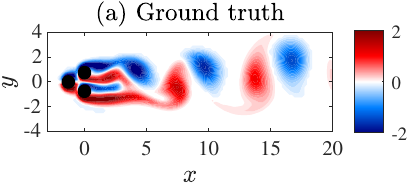}
	\includegraphics[scale=0.8]{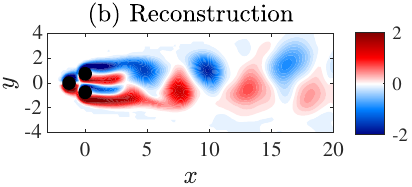}
	\includegraphics[scale=0.8]{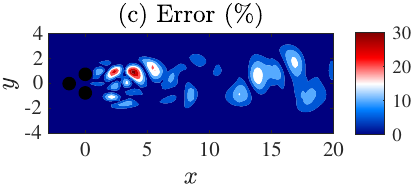}
	\caption{\label{fig:omega_contour_recon_Re_130} Snapshot of the vorticity field $\omega(x,y)$ from the high-dimensional simulation (ground truth solution, (a)) compared with the low-dimensional reconstruction by means of the three leading PDMs coordinates (b). The relative percentage error is also reported in (c). $Re=130$.}
\end{figure}

\begin{table}
	\begin{center}
		%	\vspace{0.2cm}
		%	\begin{tabular}{lcccccc}
		%		\hline
		%		\textbf{Percentile} & \textbf{RMSE} & \textbf{L1} & \textbf{L2} & \textbf{L$\infty$} \\
		%		\hline
		%5  & 1.57 & 1.32 & 1.57 & 1.74 \\
		%10 & 1.63 & 1.40 & 1.63 & 1.80 \\
		%50 & 1.93 & 1.99 & 1.93 & 2.14 \\
		%90 & 2.71 & 2.94 & 2.71 & 2.64 \\
		%95 & 2.92 & 3.16 & 2.92 & 2.87 \\
		%		\hline
		%		\textbf{Mean} & \textbf{RMSE} & \textbf{L1} & \textbf{L2} & \textbf{L$\infty$}\\
		%		\hline
		%        -- & 2.05 & 2.05 & 2.05 & 2.19 \\
		%	\end{tabular}
		\renewcommand{\arraystretch}{1.5} % Adjust the value as needed
	\begin{tabular}{lcccccc}
	\hline
	\textbf{Percentile} & 	5  & 	10 & 	50 & 	90 & 	95 &\\
	\hline
	& 9.30 \%& 10.64 \%& 18.70 \%& 31.16 \%& 35.73 \% \\
	\hline
	\textbf{Mean value} & 20.07 \%\\
	\hline
\end{tabular}
	\end{center}
	\caption{Distribution of the relative percentage error $\varepsilon_m$ between the ground truth solution and the reconstruction with three leading Diffusion Maps coordinates over the entire data-set. Reynolds number $Re=130$.}
	\label{tab:Re_130_recon_error_DM_GH}
\end{table}

The three leading parsimonious diffusion maps coordinates are employed to reconstruct the vorticity field at each time instant, following the procedure previously described in \S~\ref{subsec:recon_err}. The comparison between the snapshot ground truth solution $\omega_m(x,y)$ and the corresponding reconstruction $\tilde{\omega}_m(x,y)$ for $m = 1000$ is reported in figure~\ref{fig:omega_contour_recon_Re_130}(a)-(b). The difference between the two instantaneous vorticity fields (normalized with respect to the ground truth solution) is shown in figure~\ref{fig:omega_contour_recon_Re_130}(c), while a more comprehensive representation of the DMs-based reconstruction performance is given in Table~\ref{tab:Re_130_recon_error_DM_GH}, which reports the mean value and the percentiles (5$^{th}$, 10$^{th}$, 50$^{th}$, 90$^{th}$ and 95$^{th}$) of the relative percentage error $\varepsilon_m$ (see~(\ref{eq:recon_err}) in \S~\ref{subsec:recon_err}) evaluated over the entire data-set. In particular, it can be seen that the 95$^{th}$ percentile of the error distribution is approximately 35\%, while the mean value is $20 \%$. We also explicitly note here that the mean value of the error reduces down to $13 \%$ if the leading seven (instead of three) PDMs eigenmodes are employed in the reconstruction (namely the 1$^{st}$, 2$^{nd}$, 3$^{rd}$, 5$^{th}$, 7$^{th}$ and 13$^{th}$ mode as highlighted in figure~\ref{fig:eigenvalues_comparison_Re_130}(b)).

\subsection{Comparison with POD/PCA-based reconstruction}
\label{subsec:comparison}

\begin{figure}%[!htb]
	\centering
	\begin{tikzpicture}
	\pgfplotsset{every axis legend/.append style={at={(0.0,1.0)},anchor=north west}}
	\begin{axis}[xlabel=$Re$,ylabel=$\varepsilon_m$ ($\%$),xmin=30,xmax=130,ymin=0,ymax=50,width=13.0cm,height=4.5cm]
	\textit{\addplot[thick,black,solid,mark=o,mark size=1.5pt]
		table [x expr=\thisrowno{0}, y expr=\thisrowno{1}] {p95_Linf_var_RE_var_met.dat}; 
		\addlegendentry{POD}
	}
	\textit{\addplot[thick,red,solid,mark=o,mark size=1.5pt]
		table [x expr=\thisrowno{0}, y expr=\thisrowno{2}] {p95_Linf_var_RE_DM.dat}; 
		\addlegendentry{DMs}
	}
	%		\textit{\addplot[thick,black,dashed]
	%	table [x expr=\thisrowno{0}, y expr=10+0*\thisrowno{1}] {Files/Recontruction_error_rms_L1_L2_Linf_Re_130.dat}; 
	%}
	\end{axis}
	\end{tikzpicture}
	\caption{Relative percentage error $\varepsilon_m$ between the ground truth solution and the reconstruction with three leading modes by varying the Reynolds number $Re$: PDMs embedding (red curve); POD?PCA embedding (black curve). Note that the 95$^{th}$ percentile of the error distribution is reported.
		\label{fig:Recontruction_error_Linf_p95}
	}
\end{figure}
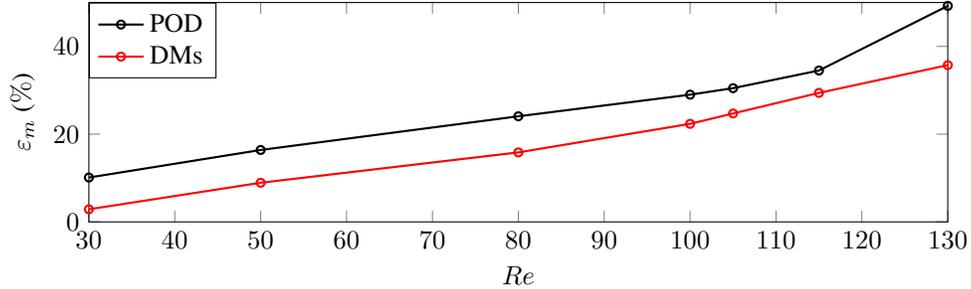

The discussion of results presented so far is concluded in this section by comparing the performance of the PDMs-based reconstruction with a counterpart linear flow reconstruction, based on the POD/PCA. This comparison is quantified in figure~\ref{fig:Recontruction_error_Linf_p95}, which reports the relative percentage error $\varepsilon_m$ (see~(\ref{eq:recon_err}) in \S~\ref{subsec:recon_err}) between the ``ground truth'' solution and the ``reconstruction'' by varying the Reynolds number $Re$. In particular, the red curve denotes the 95$^{th}$ percentile of the DMs-based error obtained with the three leading PDMs coordinates, while the black curve represents the corresponding result employing the three leading POD modes. Note that additional cases with respect to the ones so far considered (i.e. $Re=30$, $80$, $105$ and $130$) have been reported.

Since the two curves reported in figure~\ref{fig:Recontruction_error_Linf_p95} are evaluated by keeping the number of (PDMs and POD/PCA) modes fixed in the whole Reynolds number range, both the error distributions are obviously increasing functions of $Re$. However, for each $Re$, the DMs-based error (red curve) is lower than the POD-based one (black curve). This clearly demonstrates that the proposed nonlinear manifold learning algorithm  outperforms a more classic counterpart linear technique in reconstructing the rich variety of flow regimes exhibited by the fluidic pinball configuration, including chaotic conditions. Moreover, it is worth noticing that the DMs-based error curve has a nearly constant slope in the whole Reynolds number range considered, while the POD-based curve is characterized by an abrupt increase in the slope as soon as the chaotic regime is approached (see again figure~\ref{fig:Recontruction_error_Linf_p95} for $Re \approx 115$). This consideration shows us that, as expected by a robust nonlinear embedding procedure, Diffusion Maps are less sensitive to the development of nonlinear chaotic conditions with respect to the (linear) POD algorithm. 

Focusing on the chaotic regime ($Re=130$), we finally stress here that by employing the leading 7 (instead of 3) PDMs, the reconstruction error reduces down to $\varepsilon_m \approx 13 \%$ (as previously discussed in \S~\ref{subsec:results-Re_130}), while 21 leading eigenmodes are necessary to achieve the same error value with POD. This last result highlights that Diffusion Maps represent a promising tool for identifying a data-driven parsimonious reduced-order-model across all the flow regimes exhibited by the fluidic pinball configuration.

\section{Conclusions}
\label{sec:conclusions}

We have presented a parsimonious nonlinear manifold learning algorithm based on diffusion maps to identify the low-dimensional manifold  embedding of the fluidic pinball dynamics. Two-dimensional direct numerical simulations of the incompressible Navier–Stokes equations have been performed to compute the viscous wake flow behind the fluidic pinball by varying the Reynolds number $Re$. Five different flow regimes have been considered, spanning from steady symmetric ($Re < 18$) to fully chaotic ($Re > 115$) conditions. In the first step, the parsimonious set of diffusion maps coordinates is found  that encode the high-dimensional simulation data into a latent low-dimensional space. We have demonstrated that the proposed algorithm accurately and clearly determines the intrinsic dimension of the normal-form dynamics, thus providing interpretable patterns across the flow regimes. Then, we reconstructed the dynamics in the ambient space by constructing a decoder by means of Geometric Harmonics. In that manner, we were able to assess the reconstruction error and compare the performance of the proposed scheme with the traditional POD/PCA approach. The comparison analysis  has demonstrated the superiority of the proposed approach in parsimoniously representing the nonlinear dynamics up to the chaotic regime. As a matter of fact, the PDMs-based reconstruction error has been found to be significantly lower than the POD/PCA-based one across all the flow regimes. Focusing in particular on chaotic conditions ($Re=130$). Hence the proposed approach appears to be a promising altenative for identifying in a data-driven manner parsimonious reduced-order-models across all the flow regimes exhibited by the fluidic pinball configuration. Future work includes the application of the methodology for other complex fluid flows, a thorough comparison with other nonlinear manifold learning algorithms such as ISOMAP, kernel PCA, but also autoencoders and finally the construction of ROMs on latent space and their numerical bifurcation analysis.
%To the authors' knowledge, this work represents the first application of the Diffusion Maps algorithm to discover parsimonious nonlinear coordinates of a two-dimensional flow configuration such as the fluidic pinball, which is characterized by a rich variety of flow regimes including chaos.

%This paper intends to be the first step towards the identification of a fully data-driven parsimonious DMs-based reduced order surrogate model across all the flow regimes exhibited by the fluidic pinball configuration.	

%\section*{Supplementary material}
%Supplementary videos are provided along with the manuscript.

%\begin{keywords}
%	Authors should not enter keywords on the manuscript, as these must be chosen by the author during the online submission process and will then be added during the typesetting process (see \href{https://www.cambridge.org/core/journals/journal-of-fluid-mechanics/information/list-of-keywords}{Keyword PDF} for the full list).  Other classifications will be added at the same time.
%\end{keywords}

%{\bf MSC Codes }  {\it(Optional)} Please enter your MSC Codes here

%\backsection[Supplementary data]{\label{SupMat}A supplementary video is provided along with the manuscript.}

\backsection[Acknowledgements]{The numerical simulations included in the present work were performed on resources granted by CINECA under the ISCRA-C project ROMFLO. The PNRR MUR, projects PE0000013-Future Artificial Intelligence Research-FAIR \& CN0000013 CN HPC - National Centre for HPC, Big Data and Quantum Computing, Gruppo Nazionale Calcolo Scientifico-Istituto Nazionale di Alta Matematica (GNCS-INdAM) partially supported C.S. }

%\backsection[Funding]{\textcolor{red}{Insert funding if any.}}

%\backsection[Declaration of interests]{The authors report no conflict of interest.}

%\backsection[Data availability statement]{The data that support the findings of this study are openly available in [repository name] at http://doi.org/[doi], reference number [reference number].}

%\backsection[Author ORCID]{
%	
%	\noindent Alessandro Della Pia, https://orcid.org/0000-0003-2989-4397,\\
%	Dimitrios Patsatzis,
%	https://orcid.org/0000-0001-9840-0018,\\
%	Lucia Russo,
%	https://orcid.org/0000-0002-0508-3939,\\
%    Costantinos Siettos,
%    https://orcid.org/0000-0002-9568-3355.
%}
%\clearpage
%\appendix
%\section{Title of appendix (if any)}
%\label{app1}

%\clearpage
%\bibliographystyle{unsrt}
%\bibliographystyle{jfm}
%\bibliography{mybibfile}

\end{document}